\documentclass[journal]{IEEEtran}
\usepackage{cite}
\usepackage{amsmath,amssymb,amsfonts,amsthm}

\usepackage[utf8]{inputenc} 
\usepackage[T1]{fontenc}

\usepackage{graphicx}
\usepackage{textcomp}
\usepackage{xcolor,soul}
\usepackage{stmaryrd}
\usepackage{algpseudocode}
\usepackage{amsmath}
\usepackage{mathrsfs}
\usepackage{bm}
\usepackage[abs]{overpic}
\usepackage[nocomma]{optidef}

\usepackage{physics}
\usepackage{tikz}
\usepackage{mathdots}
\usepackage{yhmath}
\usepackage{cancel}
\usepackage{siunitx}
\usepackage{array}
\usepackage{multirow}
\usepackage{amssymb}
\usepackage{gensymb}
\usepackage{tabularx}
\usepackage{extarrows}
\usepackage{booktabs}
\usepackage[ruled,linesnumbered]{algorithm2e}
\makeatletter
\newcommand{\mathleft}{\@fleqntrue\@mathmargin0pt}
\newcommand{\mathcenter}{\@fleqnfalse}
\makeatother

\usetikzlibrary{fadings}
\usetikzlibrary{patterns}
\usetikzlibrary{shadows.blur}
\usetikzlibrary{shapes}

\theoremstyle{plain}
\newtheorem*{theorem*}{Theorem}

\newtheorem{remark}{Remark}
\newtheorem*{eg}{Example}

\usepackage{xcolor,varwidth}

\usepackage[scaled=.75]{beramono}
\newcommand{\bpara}[1]		{\medskip \noindent {\bf #1}}

\renewcommand\geq\geqslant
\renewcommand\leq\leqslant

\newcommand\fig[1]			{Fig.~\ref{#1}}

\def\eg					    {\emph{e.g.}~}
\def\ie					    {\emph{i.e.}~}

\usepackage{xspace}

\DeclareDocumentCommand{\sd}{o}  
{{\underline{\ast}\IfValueT{#1}{_{#1}}}}

\DeclareDocumentCommand{\xmod}{m o o o}  
{%
\IfNoValueTF{#4}
{{#1}\IfValueT{#2}{_{m,\mathsf{#2}}}\IfValueT{#3}{#3}}
{{#1}\IfValueT{#2}{_{m,\mathsf{#2}}^{\mathsf{#4}}}\IfValueT{#3}{#3}}
}

\usepackage[inline]{enumitem}
\DeclareMathAlphabet{\mathsfit}{T1}{\sfdefault}{\mddefault}{\sldefault}
\SetMathAlphabet{\mathsfit}{bold}{T1}{\sfdefault}{\bfdefault}{\sldefault}
\usepackage[switch]{lineno}
\usepackage{siunitx}
\def\BibTeX{{\mathrm B\kern-.05em{\sc i\kern-.025em b}\kern-.08em
    T\kern-.1667em\lower.7ex\hbox{E}\kern-.125emX}}
    
\usepackage{hyperref}
\usepackage{etoolbox}

\begin{document}

%\title{Max-Min Fairness Design for Energy-Efficient Quantized ISAC LEO Satellite Systems: \\ A Rate-Splitting Approach
\title{
Rate-Splitting Multiple Access for Quantized ISAC LEO Satellite Systems: A Max-Min Fair Energy-Efficient Beam Design
% Max-Min Fair Energy-Efficient Beam Design\\ for Quantized ISAC LEO Satellite Systems:\\ A Rate-Splitting Approach
}

\author{Ziang Liu,
        Longfei Yin,
        Wonjae Shin,~\IEEEmembership{Senior Member,~IEEE}
        and~Bruno Clerckx,~\IEEEmembership{Fellow,~IEEE}
\thanks{Z. Liu, L. Yin and B. Clerckx are with the Communications \& Signal Processing (CSP) Group at the Dept. of Electrical and Electronic Engg., Imperial College London, SW7 2AZ, UK. (e-mails:\{ziang.liu20, longfei.yin17, b.clerckx\}@imperial.ac.uk).

W. Shin is with the School of Electrical Engineering, Korea University, Seoul 02841, South Korea (e-mail:
wjshin@korea.ac.kr).

This work was supported in part by the National Research Foundation of Korea (NRF) grants (No.2022R1A2C4002065), and in part by the Institute of Information \& Communications Technology Planning \& Evaluation (IITP) grants (No.2024-00359235, No.2022-0-00704,  No.2021-0-00260) (\textit{Corresponding author: Wonjae Shin.)}}}
% \thanks{B. Clerckx is also with Silicon Austria Labs (SAL), Graz A-8010, Austria.}
% <-this % stops a space
% \thanks{J. Doe and J. Doe are with Anonymous University.}% <-this % stops a space

% \and
% \and
% \and
% \and
% \and

\maketitle

\begin{abstract}
Low earth orbit (LEO) satellite systems with sensing functionality are envisioned to facilitate global-coverage service and emerging applications in 6G.
% Integrated sensing and communications (ISAC)-low earth orbit (LEO) satellite communications, is envisioned as a promising technique to facilitate sensing functionality, global-coverage service in future wireless communications systems. 
Currently, two fundamental challenges, namely, inter-beam interference among users and power limitation at the LEO satellites, limit the full potential of the joint design of sensing and communication. To effectively control the interference, a rate-splitting multiple access (RSMA) scheme is employed as the interference management strategy in the system design. On the other hand, to address the limited power supply at the LEO satellites, we consider low-resolution quantization digital-to-analog converters (DACs) at the transmitter to reduce power consumption, which grows exponentially with the number of quantization bits.
Additionally, optimizing the total energy efficiency (EE) of the system is a common practice to save the power. However, this metric lacks fairness among users. To ensure this fairness and further enhance EE, we investigate the max-min fairness EE of the RSMA-assisted integrated sensing and communications (ISAC)-LEO satellite system. In this system, the satellite transmits a quantized dual-functional signal serving downlink users while detecting a target. Specifically, we optimize the precoders for maximizing the minimal EE among all users, considering the power consumption of each radio frequency (RF) chain under communication and sensing constraints. To tackle this optimization problem, we proposed an iterative algorithm based on successive convex approximation (SCA) and  Dinkelbach's method. {Numerical results illustrate that the proposed design and RSMA architecture outperforms strategies maximizing the total EE of the system, space-division multiple access (SDMA), and orthogonal multiple access (OMA) in terms of max-min fairness EE and the communication-sensing trade-off}.

% This optimization problem is first transformed to a SDP problem, and solved by 

% To enable the sensing functionality and improve the spectrum and hardware efficiency, integrated sensing and communications (ISAC)-LEO system is recently proposed. 
\end{abstract}

% \begin{IEEEkeywords}
% Energy efficiency (EE), integrated sensing and communications (ISAC), low-resolution digital-to-analog converters (DAC), MIMO radar, rate-splitting multiple access (RSMA)
% \end{IEEEkeywords}

\section{Introduction}
Sensing capabilities are recognized as a key technique in the next generation of wireless communication networks (\ie 6G) for achieving emerging applications, such as assisted navigation, activity detection and movement tracking, and environmental monitoring \cite{yazar20206g, wang2023acceleration}. Furthermore, given the significant development in the wireless communications and radar sensing industries, spectrum resources are becoming increasingly scarce and valuable. As a promising solution, integrated sensing and communication (ISAC) has been proposed to meet these requirements.
By adopting the ISAC, spectral, hardware, and energy resources are efficiently utilized. This efficiency is achieved via unified signal processing procedures and a shared hardware framework between sensing and communication systems.
% By adopting the ISAC system, spectral, hardware and energy resources are efficiently utilized due to the unified signal processing procedures, and hardware framework shared between sensing and communication systems. 
Nonetheless, much of the past ISAC research has been primarily devoted to terrestrial networks, resulting in limitations in providing global services \cite{yin2023integrated}.

On the other hand, satellite communications have been envisioned as a solution to satisfy the growing demand for ubiquitous and high-capacity global connectivity in unserved (\eg marine, and desert) and underserved areas (\eg rural areas). The deployment of satellites and terrestrial stations offers global coverage; thus, it enhances the service availability in such areas and enables service reliability through the redundancy in the satellite-terrestrial networks. In comparison to geostationary and medium earth orbit (GEO/MEO) satellites, which suffer from high propagation delay and require large investments, low earth orbit (LEO) satellites offer a lower latency, higher data rate, and near-complete visibility due to their relatively lower orbit altitudes (\ie $500-2000$ km).  Furthermore, the proximity of LEO satellites to Earth has prompted the exploration of remote sensing functionality in LEO satellite systems. In recent years, the necessity for sensing capabilities and enhanced spectrum utilization efficiency in LEO satellites has naturally driven the development of ISAC-LEO systems. However, the implementation of ISAC in LEO satellite systems poses certain challenges, potentially limiting the technological promise of ISAC-LEO systems, as follows.

\bpara{Challenge \#1: Interference Control.} 
The first challenge is the inter-beam interference among the users and the control of the trade-off between communication and sensing capabilities, as these factors determine the overall system performance. Rate-splitting multiple access (RSMA), recognized as a generalized and powerful non-orthogonal transmission framework and robust interference management strategy for multi-antenna multi-user networks, has been demonstrated to be effective in handling various interference challenges in satellite communication \cite{yin2022rate, yin2023integrated, clerckx2023primer,park2023rateten,lee2023coordinated, lyu2024rate, yin2020rate, si2022rate,yin2021rateconf, caus2018exploratory}. In RSMA, the transmitter splits each message into common and private messages, which are then linearly precoded at the satellite gateway.
% As a generalized and powerful non-orthogonal transmission framework, and robust interference management strategy for multi-antenna multi-user networks, rate-splitting multiple access (RSMA) has been proven to effectively manage different interference in satellite communication \cite{yin2022rate, yin2023integrated}. 
% Specifically, each message is split into common and private messages at the transmitter, and then linearly precoded at the satellite gateway. 
At the receiver, the common stream is first decoded by treating all private streams as noise. Subsequently, each user eliminates the decoded common stream and decodes their respective private stream, treating the other private streams as noise. Due to the ability to partially decode the interference and partially treat it as noise, RSMA outperforms the existing multiple access schemes, \eg space division multiple access (SDMA), orthogonal multiple access (OMA), and non-orthogonal multiple access (NOMA), in terms of spectral efficiency and user fairness \cite{mao2022rate}. Furthermore, RSMA has been shown to facilitate a better trade-off between the communication and sensing functionalities, when compared to conventional beamforming techniques \cite{xu2021rate, yin2022ratewcnc, yin2022rateletter, dizdar2022energy}.

\bpara{Challenge \#2: Energy Efficiency.} 
The second challenge pertains to the power limitations typically faced by LEO satellites due to the lack of its suitable power sources. This issue becomes notably more severe, especially as the number of transmit antennas increases for the improved spectral efficiency and spectrum reuse factor. Consequently, the satellite power needs to be used with high efficiency. When considering the downlink radio frequency (RF) chain, it has been shown that the digital-to-analog converter (DAC) dominates the total power consumption of the RF chain \cite{walden1999analog}.
% digital-to-analog converter (DAC) has been shown that it dominates the total power consumption of the RF-chain \cite{walden1999analog}. 
Specifically, the power consumption of DAC grows exponentially with the number of bits $b$ that determine the resolution of the DAC. By leveraging this property, we can employ low-resolution DACs, thereby reducing power consumption in each RF chain and the total power consumed at the satellite. 

Furthermore, considering the expected increase in the number of connected devices and data rates in future satellite communication systems \cite{chen2020vision}, coupled with the limitation of satellite power, it is crucial to adopt energy efficiency (EE) as a significant performance indicator in transmission scheme design. The EE is defined as the ratio between the transmission rate and the consumed energy, measured in bits/Joule. From a system-level perspective, it is preferable to consider a total EE, which is defined as the sum-rate of the communication system over the total power consumption \cite{zappone2015energy}. 
However, this metric tends to prioritize the EE in a few links with favorable channel conditions. This prioritization potentially neglects the needs of others.
% However, this metric tends to prioritize the EE in a few links with favorable channel conditions, potentially neglecting the needs of others. 
In essence, the total EE metric lacks fairness among all links, especially when dealing with a substantial number of users. Therefore, it is more reasonable to aim for maximizing the minimum EE among all links to ensure fairness.

\bpara{Related Works.}
Existing ISAC studies primarily concentrate on the terrestrial networks \cite{yin2023integrated, liu2020joint}, exploring applications such as autonomous vehicles, environment monitoring, smart cities, smart manufacturing, and industrial IoT, etc \cite{liu2022integrated}. Nevertheless, terrestrial ISAC systems fail to provide global service, and are constrained by limited data reception and processing resources. To address these challenges, the concept of an ISAC-LEO satellite system has naturally emerged, driven by the development of on-board processing capacities on the satellite. A hybrid beamforming design for ISAC-LEO system is proposed in \cite{you2022beam}, taking into account the effects of beam squint. This issue is formulated as an EE optimization, subject to the constraint of matching the beampattern. It has been demonstrated that a trade-off can be achieved between communication EE and the sensing beampattern.

\begin{table}[t]
\setlength\tabcolsep{2pt}
  \centering

\caption{Novelty Comparison with Existing RSMA-ISAC/Satellite Literature}
\resizebox{0.5\textwidth}{!}{%
  \begin{tabular}{l|c  c c c c c c c c c c}
    \toprule[1pt]
    & Our work   &  \cite{xu2021rate, loli2022rate} & \cite{yin2022ratewcnc}  & \cite{yin2022rateletter} & \cite{dizdar2022energy} & \cite{chen2023rate} & \cite{yin2020rate, si2022rate, yin2021rateconf}&  \cite{huang2022deep}  &  \cite{khan2023rate} & \cite{schroder2023comparison} & \cite{he2023secure} \\ \hline \hline
    
    ISAC                & \checkmark & \checkmark & \checkmark  & \checkmark & \checkmark & \checkmark&  &   &   &   &   \\ \hline
    
    Satellite                 & \checkmark &            & \checkmark  &            &            &  & \checkmark  &  \checkmark & \checkmark  &  \checkmark & \checkmark  \\ \hline

    Rate metric         &           & \checkmark  &             & \checkmark &            &   \checkmark & \checkmark & \checkmark & \checkmark  & \checkmark  & \checkmark  \\ \hline

    EE metric           &           &             &             &            &            \checkmark & &  &   &   &  &   \\ \hline

    Max-min EE metric   & \checkmark &  &   &   &   &           &   &   &   &   \\ \hline

    Beampattern matching&            & \checkmark &             &           &          \checkmark  & &  &   &   &   &   \\ \hline

    CRB                 & \checkmark &            & \checkmark & \checkmark &            & \checkmark & &   &   &   &   \\ \hline

    Beamforming design  & \checkmark & \checkmark & \checkmark &            & \checkmark & \checkmark & \checkmark &   &   &  & \checkmark  \\ \hline

    Power allocation    &            &            &            &            &            &       &     & \checkmark & \checkmark & \checkmark &   \\ \hline

    Low-resolution DAC  & \checkmark &            &             &            &         \checkmark  &   &   &   &   &   &   \\ \hline
    Radar sequence  & \checkmark &   \checkmark         &             &            &           & &  &   &   &   &   \\ \hline
    \bottomrule[1pt]
  \end{tabular}
  \label{tab:1}
  }
\end{table}
% RSMA is an effective interference management strategy, thus it is proposed in the system design to enhance the management of inter-beam interference among communication users, and achieve a satisfactory trade-off between communication and sensing functionalities. Starting from the RSMA-assisted ISAC system without satellite set-up, the concept of integrating RSMA with ISAC systems is initially introduced in \cite{xu2021rate}. Building upon this, \cite{loli2022rate} delves into the study of RSMA-assisted ISAC systems, considering practical scenarios that involve partial channel state information at the transmitter (CSIT) and the mobility of communication users. This investigation revealed that RSMA exhibited superior interference management capabilities, thereby improving the trade-off between the weighted sum rate (WSR) and mean square error (MSE) in beamforming approximation, when compared to conventional strategies like SDMA and NOMA. In \cite{dizdar2022energy}, a design for RSMA-assisted ISAC utilizing low-resolution DACs is investigated. This design shows the potential of RSMA to enhance EE with a reduced number of RF chains. 

RSMA is an effective interference management strategy, resulting in enhanced spectral efficiency and energy efficiency. Therefore, RSMA has been utilized in satellite systems \cite{yin2020rate, si2022rate, yin2021rateconf, huang2022deep, khan2023rate, schroder2023comparison, he2023secure} to improve communication performances. In detail, RSMA for multigroup multicast beamforming is investigated in (multigateway) multibeam satellite system \cite{yin2020rate, si2022rate, yin2021rateconf} to improve the rate performance. A deep reinforcement learning method is proposed in \cite{huang2022deep} to obtain the optimal power allocation to maximize the sum rate in satellite system. In \cite{khan2023rate}, the power allocation over different beams in satellite system is optimized by maximizing the sum rate. A comprehensive comparison of sum rate performance among RSMA, SDMA, and OMA in satellite systems is provided in \cite{schroder2023comparison}. Additionally, the secure delivery issue in investigated in \cite{he2023secure}.

RSMA is also proposed in the ISAC design to enhance the management of inter-beam interference among communication users, and achieve a satisfactory trade-off between communication and sensing functionalities. Starting from the RSMA-assisted ISAC system without satellite set-up, the concept of integrating RSMA with ISAC systems is initially introduced in \cite{xu2021rate}. Building upon this, \cite{loli2022rate} delves into the study of RSMA-assisted ISAC systems, considering practical scenarios that involve partial channel state information at the transmitter (CSIT) and the mobility of communication users. This investigation revealed that RSMA exhibited superior interference management capabilities, thereby improving the trade-off between the weighted sum rate (WSR) and mean square error (MSE) in beamforming approximation, when compared to strategies like SDMA and NOMA. In \cite{dizdar2022energy}, a design for RSMA-assisted ISAC utilizing low-resolution DACs is investigated. This design shows the potential of RSMA to enhance EE with a reduced number of RF chains. 
 
The sensing performance in the above research is optimized by the beampattern matching constraint, which solely captures the transmitter design. In contrast, Cram\'er–Rao Bound (CRB) is a theoretical lower bound on the variance of unbiased estimators, resulting in a focus on the optimization of estimation performance \cite{Kay_B_1993}. Using the CRB metric, RSMA-assisted ISAC-LEO system has been proposed recently \cite{yin2022ratewcnc, yin2022rateletter}. 
RSMA has been proven to effectively manage different types of interference and enable a better trade-off between communication and sensing in ISAC-LEO systems compared to SDMA-assisted methods.
% RSMA has been proven to be a promising approach to managing different types of interference, and enabling a better trade-off between communication sensing in ISAC-LEO system when compared to SDMA-assisted method. 
Though the extension of the multi-target case is recently investigated in \cite{chen2023rate}, it does not delve into aspects such as the EE fairness among users, low-resolution DACs, and the unique features of the satellite scenario. A novelty comparison table (cf. Table \ref{tab:1}) on existing RSMA-ISAC/Satellite is provided. In summary, previous research has not rigorously investigated the max-min fairness EE problem and interference control in the quantized ISAC-LEO satellite system, which motivates this research.

\bpara{Contributions and Overview of Results.} 
In this paper, we investigate the maximization of the minimum EE problem within the RSMA framework for an ISAC-LEO satellite system. Here, low-resolution DACs are employed in the satellite to enhance the EE. Our contributions are summarized as follows:
\begin{itemize}
    \item We propose an RSMA-assisted ISAC satellite system design to facilitate multi-user communications and target sensing, simultaneously. Considering the non-linearity of the quantization effect in low-resolution DACs, we define a max-min fairness EE to ensure equitable EE performance across all users. We adopt the RSMA transmission scheme to enable effective interference control, and achieve an optimized communication-sensing trade-off. Our proposed design addresses two challenges: interference control and fairness EE guarantee among all users in the context of ISAC-LEO systems. The latter is normally neglected as most previous studies adopt the total EE metric. We show that our proposed design has a higher max-min fairness EE compared to design by total EE metric. It is observed that adopting low-resolution DACs effectively improves max-min fairness EE. Specifically, it is enhanced with satisfactory detection performances (\ie Doppler, and range), when the quantization bit decreases. Compared to previous RSMA ISAC studies \cite{xu2021rate, yin2022ratewcnc, yin2022rateletter, dizdar2022energy,loli2022rate, chen2023rate}, the consistent observation is that the radar sequence is not utilized when maximizing sum-rate metric. Consequently, a common stream is used for both communication and sensing functions. However, the new observation is that, when adopting max-min fairness EE as the metric, the radar sequence is used for sensing. This is due to the utilization of a separated radar sequence facilitating a higher EE; the power of the radar sequence is used for sensing and the power consumed by common and private streams, which determines EE, is minimized mainly for communication use.

    \item We formulate the maximization of the minimum EE as a non-convex optimization problem to obtain the optimal precoders. The formulated problem is subject to total transmit power, communication and sensing constraints. We ensure communication performance through quality-of-service (QoS) requirements while maintaining sensing estimation performance by considering the CRB on angle estimates. We transform the problem into a semi-definite programming (SDP) problem. Optimal precoders can be then obtained by solving the SDP problem through an iterative algorithm based on successive convex approximation (SCA) and Dinkelbach's method.

    \item The numerical results are provided to illustrate the performance of the proposed RSMA-assisted ISAC-LEO satellite architecture. Key observations include: (a) {The proposed design outperforms the strategies maximizing the total EE of the system, SDMA and OMA in terms of max-min fairness EE}; (b) RSMA facilities an improved trade-off between communication (\ie max-min fairness EE) and sensing functionalities; (c) RSMA, combined with an additional precoded radar sequence with SIC operation, can lead to the best max-min fairness EE;
    % RSMA with an additional precoded radar sequence with SIC operation can lead to the best max-min fairness EE; 
    (d) A lower DAC quantization bit results in a higher max-min fairness EE with satisfactory detection performances.

\end{itemize}

\bpara{Organization of This Paper.} The paper organization is as follows. We revisit the system model in Section II. In Section III, the metrics and problem formulation are introduced. The problem transformation, the proposed solution algorithm, and a brief computational complexity and convergence analysis are provided in In Section IV. Section V provides numerical evaluations, and we conclude this work in Section VI.

\bpara{Notation.} The set of integers, real numbers, and complex numbers are respectively represented by $\mathbb{Z}$, $\mathbb{R}$,  and $\mathbb{C}$. 
Continuous signals and discrete sequences are denoted by $x(t), t \in \mathbb{R}$ and $x[k], k \in \mathbb{Z}$, respectively. Matrices, vectors and scalars are expressed by capital boldface, small boldface and normal fonts, respectively. 
We denote conjugate-transpose and transpose of matrix $\mathbf{X}$ by $\mathbf{X}^H$ and $\mathbf{X}^\top$, respectively. $|\cdot|$, $\mathbb{E}{(\cdot)}$, $\|\cdot\|_{2}$, and $\circ$ are utilized to denote absolute value, statistical expectation, Euclidean norm, and Hadamard product.

\section{System model}
\begin{figure}[tb]
\centering
		\includegraphics[width = 0.4\textwidth]{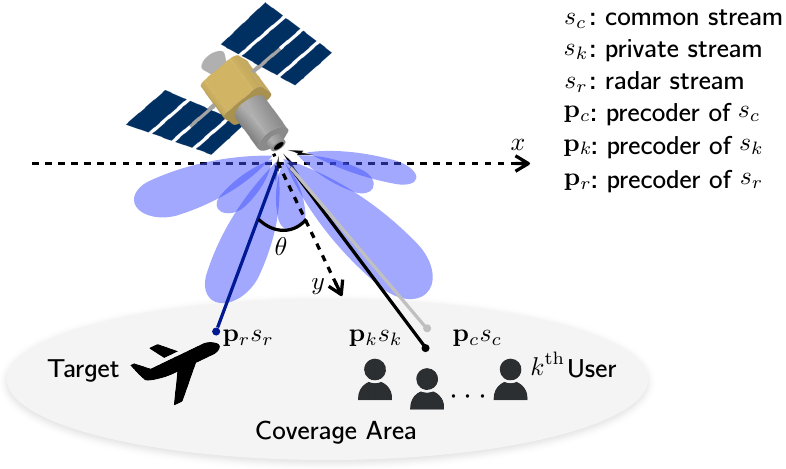}
		\caption{Integrated sensing and communications enabled quantized rate-splitting multiple access satellite system.}
		\label{fig:system_model}
\end{figure}

As depicted in \fig{fig:system_model}, we consider a downlink ISAC-LEO satellite system equipped with co-located $N_t$ transmit antennas and $N_r$ receive antennas (\ie monostatic setup). 
{This system operates in full-duplex mode to prevent echo-miss problem, which arises from signal transmission durations being considerably longer than the radar echo round-trip time \cite{liu2023joint}.
% due to the signal transmission durations are typically much longer than the round-trip times for radar echo \cite{liu2023joint}.} 
This system serves $K$ single-antenna users, indexed by $\mathcal{K} = \{1, \cdots, K\}$, and detects a target. We assume that all antenna arrays are uniform linear arrays (ULA)\cite{roper2022beamspace}, featuring half-wavelength spacing between adjacent antenna elements. As shown in \fig{fig:rsma_tx}, we adopt the RSMA scheme based on the 1-layer RS originally proposed in \cite{clerckx2016rate, joudeh2016sum}. Thus, the message $W_k$ for the $k^{\text{th}}$ user is categorized into a common message $W_{c,k}$ and a private message $W_{p,k}$. Subsequently, all common messages, $\{ W_{c,1}, \cdots ,W_{c,K} \}$, are jointly encoded into a common stream $s_c$. All the private parts, $\{ W_{p,1}, \cdots ,W_{p,K} \}$, are encoded into private streams,  $\{ s_{1}, \cdots ,s_{K} \}$. The transmit streams comprise common stream, private streams, and radar sequence are expressed as $\mathbf{s}[\ell] = [s_c[\ell], s_1[\ell], \cdots, s_K[\ell], s_r[\ell]]^\top \in \mathbb{C}^{(K+2) \times 1}$, where $\ell \in \mathcal{L} = \{1, \cdots, L\}$ is the discrete time index.
% We can denote the transmit streams including the radar sequence as $\mathbf{s}[\ell] = [s_c[\ell], s_1[\ell], \cdots, s_K[\ell], s_r[\ell]]^\top \in \mathbb{C}^{(K+2) \times 1}$, where $\ell \in \mathcal{L} = \{1, \cdots, L\}$ is the discrete time index. 
Specifically, the radar sequence is deterministic and known at the receiver for target information extraction. 
\begin{figure}[tb]
\centering
		\includegraphics[width = 0.45\textwidth]{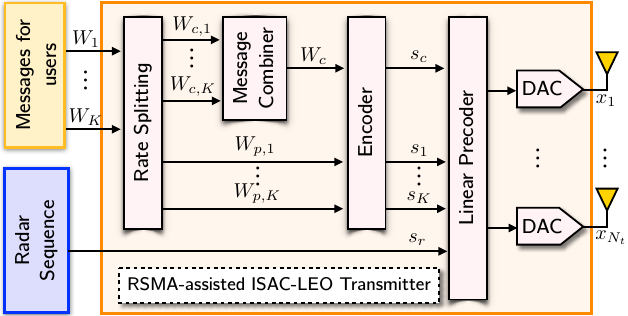}
		\caption{Transmitter Architecture of RSMA-assisted ISAC-LEO system.}
		\label{fig:rsma_tx}
\end{figure}
\subsection{Channel Model}
\bpara{Satellite Communication Channel Model.} The satellite channel comprises free space loss, radiation pattern and rain attenuation, thus the downlink channel between the LEO ISAC satellite and the $k^{\text{th}}$ user is modelled as \cite{yin2022rate, Zheng_J_2012}
\begin{equation}
\label{eq:satelliteh}
\mathbf{h}_{k}=\mathbf{b}_{k} \circ \mathbf{q}_{k} \circ \exp \left\{\jmath \boldsymbol{\phi}_{k}\right\},
\end{equation}
where $\mathbf{h}_k \in \mathbb{C}^{N_t \times 1}$ is the downlink channel between the $k^{\text{th}}$ user and the LEO satellite based on the satellite setup \cite{gharanjik2015robust}, $\mathbf{b}_{k} = [b_{k,1}, b_{k,2}, \cdots, b_{k,N_t}] \in \mathbb{C}^{N_t}$ consists of satellite beam radiation pattern and free space loss. The $n_t^{\text{th}}$ element of the $\mathbf{b}_{k}$ is approximated as 
\begin{equation}
b_{k, n_t}=\frac{\sqrt{G_\mathsf{u} G_{k, n_t}}}{4 \pi \frac{d_{k}}{\lambda_\mathsf{c}} \sqrt{\iota T_\mathsf{sys} B}},
\end{equation}
where $G_\mathsf{u}$ is the antenna gain at the user, $d_{k}$ is the distance between the satellite and the $k^{\text{th}}$ user, $\lambda_\mathsf{c}$ is the carrier wavelength, $\iota$ is the Boltzman’s constant, $T_\mathsf{sys}$ is the receive noise temperature, $B$ is the bandwidth. $G_{k, n_t}$ is the beam gain from the $n_t^{\text{th}}$ feed to the $k^{\text{th}}$ user, and is approximated by
\begin{equation}
    G_{k, n_t}=G_{\mathsf{max}}\left[\frac{J_1\left(u_{k, n_t}\right)}{2 u_{k, n_t}}+36 \frac{J_3\left(u_{k, n_t}\right)}{u_{k, n_t}^3}\right]^2,
\end{equation}
where $G_{\mathsf{max}}$ denotes the maximum beam gain for each beam and $u_{k, n_t}=2.07123 \sin \left(\theta_{k, n_t}\right) / \sin \left(\theta_{3 \text{ dB}}\right)$. $\theta_{k, n_t}$ is the angle between the $k^{\text{th}}$ user and the center of $n_t^{\text{th}}$ beam, and $\theta_{3 \text{ dB}}$ is the $3$  dB angle loss compared with the gain of the beam center. $J_1$ and $J_3$ are the first and third order of the first-kind Bessel function, respectively. $\mathbf{q}_{k} = [q_{k,1}, q_{k,2}, \cdots, q_{k,N_t}] \in \mathbb{C}^{N_t}$ is the rain attenuation coefficient vector, where each element ${q}_{k,n_t} = \zeta^{1/2}_{k,n_t}$. The power gain $\zeta_{k,n_t}$ in  dB, $\zeta_{k,n_t}(\mathsf{ dB}) = 20\log_{10}(\zeta_{k,n_t})$, commonly follows the lognormal distribution, \ie $\ln{(\zeta_{k,n_t}(\mathsf{ dB}))} \sim \mathcal{N}(\mu, \sigma)$. Additionally, $\boldsymbol{\phi}_k = [{\phi}_{k,1}, {\phi}_{k,2}, \cdots, {\phi}_{k,N_t}] \in \mathbb{C}^{N_t}$ is the uniformly distributed phase vector, where ${\phi}_{k,n_t} \sim \mathcal{U}(0, 2 \pi)$. We denote the satellite channels between the satellite and all the users by $\mathbf{H} = [\mathbf{h}_1, \cdots, \mathbf{h}_K] \in \mathbb{C}^{N_t \times K}$.

\bpara{Radar Channel Model.} 
We consider a radar operating in a track model, with an initial Angle-of-Arrival (AoA) estimate for the target obtained from prior scanning.
By processing the received reflected echo, we can extract the information of the target, \ie target’s AoA, $\theta$, range, $r$, and  {the relative velocity, $v$, between the LEO satellite and target in the line-of-sight (LOS) direction}. Since the AoA and Angle-of-Departure (AoD) are the same for the monostatic radar, they are both denoted as $\theta$. We assume that the radar channel varies slowly with time, thus it is modeled by \cite{liu2023joint}
\begin{equation}
    \mathbf{H}_{r}(t) = \alpha e^{\jmath 2 \pi f_d \ell T} \mathbf{b}(\theta) \mathbf{a}^\top(\theta),
\end{equation}
\begin{enumerate}[leftmargin = *, label = --]
	\item $\alpha$ is the attenuation factor, comprising radar cross-section (RCS) and round-trip path-loss; $\nu$ is derived from the radar range equation \cite{skolnik1980introduction}, given in \eqref{eq:radar_fad}; $r$ is the distance between the receiver and the target.
    \begin{equation}
        \alpha = \sqrt{\frac{\lambda_c^{2} \nu}{(4 \pi)^{3} r^{4}}}.
        \label{eq:radar_fad}
    \end{equation}
    
    \item $\ell$ is the discrete time index; $T$ is the sampling interval; $t = \ell T$ is the time instant; $f_d = 2 v f_c /c$ is the Doppler shift; the signal carrier frequency and the speed of light are denoted by $f_c$ and $c$, respectively.
    
    \item Transmit and receive steering vectors of the ULA, $\mathbf{a}(\theta) \in \mathbb{C}^{N_t \times 1}$ and $\mathbf{b}(\theta) \in \mathbb{C}^{N_r \times 1}$, are respectively written as
    \begin{subequations}
    \begin{align}
        \mathbf{a}(\theta)  & = \left[1, e^{\jmath 2 \pi \omega \sin \left(\theta\right)}, \ldots, e^{\jmath 2 \pi\left(N_{t}-1\right) \omega \sin \left(\theta\right)}\right]^\top, \\
        \mathbf{b}(\theta) & = \left[1, e^{\jmath 2 \pi \omega \sin \left(\theta\right)}, \ldots, e^{\jmath 2 \pi\left(N_{r}-1\right) \omega \sin \left(\theta\right)}\right]^\top, 
    \end{align}
    \end{subequations}
    \end{enumerate}
 % where $\omega$ is the normalized distance between adjacent array elements relative to wavelength.
where $\omega$ represents the normalized spacing between adjacent array elements in relation to the wavelength.

\subsection{Signal Model}
The discrete baseband dual-functional signal at time index $\ell \in \mathcal{L}$ is as follows
% \begin{equation}
% \begin{aligned}
% \mathbf{x}[\ell] & = \mathbf{P} \mathbf{s} [\ell]\\
% & =  \underbrace{\overbrace{\mathbf{p}_{c} s_{c}[\ell] \vphantom{\sum}}^{\text {Common stream }}+
% \overbrace{\sum_{k \in \mathcal{K}} \mathbf{p}_k s_k[\ell]}^{\text {Private streams }}}_{\text {Communication streams }}+
% \underbrace{\mathbf{p}_{r} s_{r}[\ell] \vphantom{\sum_{k \in \mathcal{K}}}}_{\text {Radar sequence }},
% \end{aligned}
% \end{equation}
\begin{equation}
\mathbf{x}[\ell] \! = \! \mathbf{P} \mathbf{s} [\ell]
\, = \!  \underbrace{\overbrace{\mathbf{p}_{c} s_{c}[\ell] \vphantom{\sum}}^{\text {Common stream }} \! + \!
\overbrace{\sum_{k \in \mathcal{K}} \mathbf{p}_k s_k[\ell]}^{\text {Private streams }}}_{\text {Communication streams }} \, + \!
\underbrace{\mathbf{p}_{r} s_{r}[\ell] \vphantom{\sum_{k \in \mathcal{K}}}}_{\text {Radar sequence }},
\end{equation}
where $\mathbf{P} = [\mathbf{p}_{c}, \mathbf{p}_{1}, \cdots, \mathbf{p}_{K}, \mathbf{p}_{r}] \in \mathbb{C}^{N_t \times (K+2)}$ is the dual-functional beamforming matrix to be optimized. Specifically, $\mathbf{p}_{c} \in \mathbb{C}^{N_t}$ is the precoder of the common stream, $\{\mathbf{p}_k\}_{k \in \mathcal{K}}$ is the precoder of the private stream, and $\mathbf{p}_{r} \in \mathbb{C}^{N_t}$ is the precoder of the radar sensing sequence. {Note that the radar sequence, $s_r$, is deterministic for extracting the target's information, and it has no useful data for the users.}

Before the transmission, the baseband signal, $\mathbf{x}[\ell]$, is quantized by the DAC. The quantization is a non-linear mapping, thus the additive quantization noise model (AQNM) \cite{orhan2015low, dizdar2022energy, park2023rate} is adopted to approximate the quantization to a linear form.
Therefore, the analog transmit signal at time $\ell$ after quantization is given by
\begin{equation}
    \mathscr{Q}(\mathbf{x}) \approx \mathbf{x}_{\mathsf{q}}[\ell] = \boldsymbol{\Delta} \mathbf{x}[\ell] + \boldsymbol{\epsilon}[\ell],
\end{equation}
where $\mathbf{x}_{\mathsf{q}}[\ell] = \mathbf{x}_{\mathsf{q},c}[\ell] + \sum_{k \in \mathcal{K}}\mathbf{x}_{\mathsf{q},k}[\ell]  + \mathbf{x}_{\mathsf{q},r}[\ell] \in \mathbb{R}^{N_t \times 1}$. The $\boldsymbol{\Delta} \triangleq \text{diag}(\delta_1, \cdots, \delta_{N_t}) \in \mathbb{R}^{N_t \times N_t}$,  {where $\delta_{m} \in (0,1)$ is the linear gain of the quantization at the $m^{\text{th}}$ transmit RF chain when adopting $b_m$ bit DAC quantization, written as
\begin{equation}
    \label{eq:delta}
    \delta_{m} = 1- \varrho = 1 - \frac{\pi \sqrt{3}}{2} 2^{-2b_m},
\end{equation}
where $\varrho$ is the quantization loss.
Additionally, $\boldsymbol{\epsilon}$ is the additive quantization noise vector,} $\boldsymbol{\epsilon} \sim \mathcal{CN}(\mathbf{0}_{N_t}, \boldsymbol{\Sigma})$, where 
\begin{equation}
    \boldsymbol{\Sigma} = \boldsymbol{\Delta} (\mathbf{I}_{N_t} - \boldsymbol{\Delta}) \mathbb{E}\big[\mathbf{x}[\ell] \mathbf{x}^H[\ell]\big].
\end{equation}
The coherent matrix of transmitted dual-functional waveform after quantization is as follows
\begin{equation}
    \mathbf{R}_\mathsf{q} = \mathbb{E}\big[\mathbf{x}_{\mathsf{q}}[\ell] \mathbf{x}^H_{\mathsf{q}}[\ell]\big] = \boldsymbol{\Delta} \mathbf{P} \mathbf{P}^H \boldsymbol{\Delta}^H.
\end{equation}

% $ \boldsymbol{\Sigma} = \text{diag} (\sigma^2_{\epsilon,1}, \cdots, \sigma^2_{\epsilon,m})$, and $\sigma^2_{\epsilon,m}) = (1 - \delta_m) \delta_m \mathbb{E}[\mathbf{x} \mathbf{x}^H]$.

The received signal at all $K$ users is given by
\begin{equation}
    \mathbf{y}[\ell] = \mathbf{H}^H \mathbf{x}_\mathsf{q}[\ell] + \mathbf{n}[\ell],
\end{equation}
where $\mathbf{n} \sim \mathcal{CN}(\mathbf{0}_{K}, \sigma^2_{k}\mathbf{I}_K)$ is the additive complex Gaussian noise at each user. We assume that the channel state information (CSI) is perfectly known at the satellite. At the $k^{\text{th}}$ user, the received signal is written as
\begin{equation}
\begin{aligned}
y_k[\ell] & =\mathbf{h}_k^H \mathbf{x}_\mathsf{q}[\ell]+n_k[\ell] \\
& =\mathbf{h}_k^H \boldsymbol{\Delta} \mathbf{p}_c s_c[\ell]+\mathbf{h}_k^H \boldsymbol{\Delta} \sum_{k \in \mathcal{K}} \mathbf{p}_k s_k[\ell]  \\
& \quad + \mathbf{h}_k^H \boldsymbol{\Delta} \mathbf{p}_r s_r[\ell]  + \mathbf{h}_k^H \boldsymbol{\epsilon}[\ell]   +n_k[\ell]. 
% \, \forall k \in \mathcal{K} .
\end{aligned}
\end{equation}
% where $\mathbf{h}_k \in \mathbb{C}^{N_t \times 1}$ is the downlink channel between the $k^{\text{th}}$ user and the LEO satellite based on the satellite \cite{gharanjik2015robust}, and $n_k[\ell] \sim \mathcal{CN} (0, \sigma^2_{n,k})$ is the additive Gaussian noise at the $k^{\text{th}}$ user. 
At ISAC-LEO satellite receiver, the radar echo, $\mathbf{z}[\ell] \in \mathbb{C}^{N_r \times 1}$, for the sensing functionality is given by 
\begin{equation}
\label{eq:radarsignal}
\begin{aligned}
\mathbf{z}[\ell] & =\mathbf{H}_{r} \mathbf{x}_\mathsf{q}[\ell]+\mathbf{n}_r[\ell] \\
& =\alpha e^{\jmath 2 \pi f_d \ell T} \mathbf{b}(\theta) \mathbf{a}^\top(\theta) \boldsymbol{\Delta}  \mathbf{x}[\ell] \\
& \quad + \alpha e^{\jmath 2 \pi f_d \ell T} \mathbf{b}(\theta) \mathbf{a}^\top(\theta) \boldsymbol{\epsilon}[\ell] + \mathbf{n}_r[\ell],
\end{aligned}
\end{equation}
where $\mathbf{n}_r[\ell]$ is the independent, zero-mean complex Gaussian noise, and $\mathbf{n}_r[\ell] \sim \mathcal{CN}(\mathbf{0}_{N_r}, \sigma^2_r \mathbf{I}_{N_r})$.

\section{Minimal Energy Efficiency Maximization and Problem Formulation}
In this section, the metrics in the objective function and constraints for an optimization problem of interest are explained. Additionally, we formulate the optimization problem.

\subsection{Communication Metric} 
\bpara{Communication Rates.} Considering the EE of each user, we adopt the max-min fairness EE. Since RSMA is adopted, the communication metric needs to be modified due to the common stream. Recalling the decoding process of RSMA \cite{mao2022rate}, each user first decodes the common stream, $s_c$, by treating all private streams as noise. After the common stream is successfully decoded, it is subtracted from the received signal, $\mathbf{y}$, by SIC. In RSMA, the interference is partially subtracted and the remaining interference is treated as noise, resulting in significant improvement of sum-rate \cite{mao2018rate} or minimum rate among all users \cite{lee2022max}. This is facilitated by the common and private message split and SIC in the decoding.

The signal-to-interference-plus-noise (SINRs) for the common and private streams at $k^{\text{th}}$ user with the assumption that channels vary slowly with time are respectively written as
\begin{subequations}
\begin{equation}
\gamma_{c, k}=\frac{\left|\mathbf{h}_k^H \boldsymbol{\Delta} \mathbf{p}_c\right|^2}{\sum_{i \in \mathcal{K}}\left|\mathbf{h}_k^H \boldsymbol{\Delta} \mathbf{p}_i\right|^2 + \psi \left|\mathbf{h}_k^H \boldsymbol{\Delta} \mathbf{p}_r\right|^2 + \mathbf{h}_k^H \boldsymbol{\Sigma} \mathbf{h}_k + \sigma_k^2},
\label{eq:sinr1}
\end{equation}
\begin{equation}
\gamma_k=\frac{\left|\mathbf{h}_k^H \boldsymbol{\Delta} \mathbf{p}_k\right|^2}{\sum_{i \in \mathcal{K}, i \neq k}\left|\mathbf{h}_k^H \boldsymbol{\Delta} \mathbf{p}_i\right|^2 + \psi \left|\mathbf{h}_k^H \boldsymbol{\Delta} \mathbf{p}_r\right|^2 + \mathbf{h}_k^H \boldsymbol{\Sigma} \mathbf{h}_k + \sigma_k^2},
\label{eq:sinr2}
\end{equation}
\end{subequations}
where $\psi$ is an index variable to unify different processing methods for radar sequence, $s_r$. Specifically, 
% two cases are considered:
we consider:
\begin{enumerate}[leftmargin = *, label = --]
    \item {$\psi = 0$: Since the radar sequence, $s_r$, is commonly designed and determined before transmission, the user side can access to the radar sequence and implement SIC to remove this interference. This case is hence denoted by $\psi = 0$.}
    \item {$\psi=1$: SIC at the user receiver can be computationally intensive due to the iterative interference cancellation, thus to facilitate low computational and hardware complexity at the user side, we can adopt no SIC at the expense of treating radar sequence, $s_r$, as the interference.}
\end{enumerate}
The rates of the common and private streams with regard to SINR are written as
\begin{subequations}
    \begin{equation}
        R_{c, k}=\log _2\left(1+\gamma_{c, k}\right),
    \end{equation}
    \begin{equation}
        R_{k}=\log _2\left(1+\gamma_{k}\right).
    \end{equation}
\end{subequations}
In the RSMA framework, the common stream needs to be decodable by all receivers. To guarantee this, the common rate, denoted by $R_{c}$, should set to the minimum of the $R_{c, k}, \forall k \in \mathcal{K}$ \cite{mao2022rate}, \ie
\begin{equation}
    R_c = \min_{k \in \mathcal{K}} \{R_{c, k} \} = \sum_{k \in \mathcal{K}} C_k,
\end{equation}
where $C_k$ is the rate of the common part of the $k^{\text{th}}$ user. Thus, the total achievable rate of the $k^{\text{th}}$ user is $R_{k, \mathsf{total}} = C_k + R_k$.

\bpara{Max-min Energy Efficiency.} Communication rates are determined on the $K$ user end, making them related to the transmission design. Thus, we consider the power consumption of the transmitter as the denominator in the max-min EE metric. In each RF chain, the power consumption comes from the low-pass filter, mixer, local oscillator, $90^\circ$ hybrid with buffer, DAC, whose power consumption are denoted by $P_\mathsf{LP}, P_\mathsf{M}, P_\mathsf{LO}, P_\mathsf{H}, P_\mathsf{DAC}$, respectively. Specifically, the power consumption of one RF chain is $P_\mathsf{RF} = 2 P_\mathsf{DAC} + 2 P_\mathsf{LP} + 2 P_\mathsf{M} + P_\mathsf{H} + P_\mathsf{LO}/N_t$, where all transmit RF chains share one local oscillator, thus $P_\mathsf{LO}$ is divided by $N_t$ for simplicity while we assume that all RF chains adopt the same quantization (\ie $b_m = b, \forall m \in \mathcal{N}_t = \{1, \cdots, N_t\}$) The DAC power consumption $P_\mathsf{DAC}$ in Watt is as follows \cite{ribeiro2018energy, choi2022energy}
\begin{equation}
P_{\mathsf{DAC}}\left(b, f_s\right)=1.5 \times 10^{-5} \cdot 2^{b}+9 \times 10^{-12} \cdot f_s \cdot b,
\end{equation}
where $f_s$ is the sampling rate. Subsequently, the total power consumption for each user is written as \cite{ribeiro2018energy}
\begin{equation}
    P_{\mathsf{chain}, k} = P_\mathsf{RF} + \kappa^{-1} P_{\mathsf{TX}, k},
\end{equation}
where $\kappa$ is the power amplifier (PA) efficiency (\ie $\kappa = P_{\mathsf{TX}, k} / P_\mathsf{PA}$), $P_{\mathsf{TX}, k}$ is the actual transmit power for the $k^\text{th}$ user, given by
\begin{equation}
    P_{\mathsf{TX}, k} = \text{tr}(\mathbb{E}[\mathbf{x}^{}_{\mathsf{q},k}\mathbf{x}^H_{\mathsf{q},k}])  
    % & = \text{tr} (\boldsymbol{\Delta} \mathbf{p}_c\mathbf{p}^H_c \boldsymbol{\Delta} + \boldsymbol{\Delta} \mathbf{p}_r\mathbf{p}^H_r \boldsymbol{\Delta} +
    % \boldsymbol{\Delta} \sum_{k\in \mathcal{K}} \mathbf{p}_k\mathbf{p}^H_k \boldsymbol{\Delta} ).
     = \norm{\boldsymbol{\Delta} \mathbf{p}_k}^2 + 1/K \norm{\boldsymbol{\Delta} \mathbf{p}_c}^2,
\end{equation}
{where we consider the power of the common stream for each user by dividing user number $K$. Therefore, the EE of the $k^{\text{th}}$ user in the considered system is as follows
% \begin{equation}
% \begin{aligned}
%     \eta_k & = \frac{R_{k, \mathsf{total}}}{P_{\mathsf{chain}, k}}\\
%     & = \frac{C_k + R_k}{P_\mathsf{RF} + \kappa^{-1} \mathbf{p}_k\mathbf{p}^H_k}
% \end{aligned}
% \end{equation}
\begin{equation}
    \eta_k = \frac{R_{k, \mathsf{total}}}{P_{\mathsf{chain}, k}}
     = \frac{C_k + R_k}{P_\mathsf{RF} + \kappa^{-1} (\norm{\boldsymbol{\Delta} \mathbf{p}_k}^2 + 1/K \norm{\boldsymbol{\Delta} \mathbf{p}_c}^2)}.
\end{equation}
}

\subsection{Sensing Metric: Cram\'er–Rao Bound}
% \begin{figure*}
% \begin{equation}
% \label{eq:crb}
% \begin{aligned}
% \mathsf{C R B}(\theta) & =\left[J_{\theta \theta}-J_{\theta {\alpha}} J_{{\alpha} {\alpha}}^{-1} J_{\theta {\alpha}}^\top\right]^{-1} \\
% & =\frac{\operatorname{tr}\left(\mathbf{A}(\theta) \mathbf{R}_{\mathsf{q}} \mathbf{A}^H(\theta)\right)}{2 \, \mathsf{S N R}_\mathsf{radar} \left(\operatorname{tr}\left(\dot{\mathbf{A}}(\theta) \mathbf{R}_{\mathsf{q}} \dot{\mathbf{A}}^H(\theta)\right) \operatorname{tr}\left(\mathbf{A}(\theta) \mathbf{R}_{\mathsf{q}} \mathbf{A}^H(\theta)\right)-\left|\operatorname{tr}\left(\mathbf{A}(\theta) \mathbf{R}_{\mathsf{q}} \dot{\mathbf{A}}^H(\theta)\right)\right|^2\right)},
% \end{aligned}
% \end{equation}
% \hrulefill
% \end{figure*}
\begin{figure*}
\begin{equation}
\label{eq:crb}
\mathsf{C R B}(\theta)  =\left[J_{\theta \theta}-J_{\theta {\alpha}} J_{{\alpha} {\alpha}}^{-1} J_{\theta {\alpha}}^\top\right]^{-1} 
 =\frac{\operatorname{tr}\left(\mathbf{A}(\theta) \mathbf{R}_{\mathsf{q}} \mathbf{A}^H(\theta)\right)}{2 \, \mathsf{S N R}_\mathsf{radar} \left(\operatorname{tr}\left(\dot{\mathbf{A}}(\theta) \mathbf{R}_{\mathsf{q}} \dot{\mathbf{A}}^H(\theta)\right) \operatorname{tr}\left(\mathbf{A}(\theta) \mathbf{R}_{\mathsf{q}} \mathbf{A}^H(\theta)\right)-\left|\operatorname{tr}\left(\mathbf{A}(\theta) \mathbf{R}_{\mathsf{q}} \dot{\mathbf{A}}^H(\theta)\right)\right|^2\right)},
\end{equation}
\hrulefill
\end{figure*}
For a target sensing, the target information extraction can be formed as a classic maximum likelihood estimation (MLE) problem \cite{Kay_B_1993}. After the estimated parameters are obtained, the mean square error (MSE) is used as the sensing performance metric. However, MSE has no close-form expression, thus intractable. Instead, the Cram\'er–Rao Bound is employed as the performance metric, which provides a lower bound for the MSE and has a closed-form expression. The CRB is calculated as $\mathsf{CRB} = \mathbf{J}^{-1}$, where $\mathbf{J}$ is the Fisher information matrix (FIM). The FIM \cite{Kay_B_1993} for estimating the unknown parameter vector $\boldsymbol{\xi} = [\theta, \alpha]^\top$ from \eqref{eq:radarsignal}, is written as
\begin{equation}
[\mathbf{J}]_{i, j}=\frac{2}{\sigma_r^2} \operatorname{Re}\left\{\sum_{\ell=1}^L \frac{\partial \boldsymbol{\mu}[\ell]^H}{\partial \xi_i} \frac{\partial \boldsymbol{\mu}[\ell]}{\partial \xi_j}\right\}, i, j \in \{1, 2\},
\end{equation}
where $\boldsymbol{\mu} \triangleq \alpha e^{\jmath 2 \pi f_d \ell T} \mathbf{b}(\theta) \mathbf{a}^\top(\theta) \boldsymbol{\Delta}  \mathbf{x}[\ell]$. Therefore, the FIM for estimation is partitioned as
\begin{equation}
\mathbf{J}_{\boldsymbol{\xi}}=\left[\begin{array}{ll}
J_{\theta \theta} & J_{\theta \alpha} \\
J_{\theta \alpha}^\top & J_{\alpha \alpha}
\end{array}\right].
\end{equation}
The CRB matrix for estimating AoA, $\theta$, is written in \eqref{eq:crb} \cite{bekkerman2006target, yin2022rateletter}
%\begin{figure*}
%\begin{equation}
%\label{eq:crb}
%\begin{aligned}
%\mathsf{C R B}(\theta) & =\left[J_{\theta \theta}-J_{\theta {\alpha}} J_{{\alpha} {\alpha}}^{-1} J_{\theta {\alpha}}^\top\right]^{-1} \\
%& =\frac{\operatorname{tr}\left(\mathbf{A}(\theta) \mathbf{R}_{\mathsf{q}} \mathbf{A}^H(\theta)\right)}{2 \, \mathsf{S N R}\left(\operatorname{tr}\left(\dot{\mathbf{A}}(\theta) \mathbf{R}_{\mathsf{q}} \dot{\mathbf{A}}^H(\theta)\right) \operatorname{tr}\left(\mathbf{A}(\theta) \mathbf{R}_{\mathsf{q}} \mathbf{A}^H(\theta)\right)-\left|\operatorname{tr}\left(\mathbf{A}(\theta) \mathbf{R}_{\mathsf{q}} \dot{\mathbf{A}}^H(\theta)\right)\right|^2\right)},
%\end{aligned}
%\end{equation}
%\hrulefill
%\end{figure*}
where 
% $\mathbf{A}(\theta) \triangleq \mathbf{b}(\theta) \mathbf{a}^\top(\theta)$, $\mathsf{S N R} = L |\alpha |^2 / \sigma^2_r$
$\mathbf{A}(\theta) \triangleq \mathbf{b}(\theta) \mathbf{a}^\top(\theta)$; $\dot{\mathbf{A}} \triangleq {\partial \mathbf{A}(\theta)}/{\partial \theta}$; the SNR of received radar echo $\mathsf{S N R}_\mathsf{radar} = P_t L |\alpha |^2 / \sigma^2_r$;  {$P_t$ is the transmitted power; $L$ represents the number of transmit symbols within one coherent processing interval (CPI), which facilities the multi-pulse processing and improves the received radar SNR \cite{richards2010principles}.} The expression of the elements in FIM (\ie $J_{\theta \theta}, J_{\theta {\alpha}}, J_{{\alpha} {\alpha}}$) are derived in Appendix A. {Note that constraining the CRB on angle effectively limits the minimum achievable error in estimating the AoA. Thus, it indirectly necessitates the transmit beamforming pattern to concentrate its power on the target. This, in turn, leads to an overall enhancement of detection performance, such as Doppler detection.}
In practice, the AoA of the radar echo, $\theta$, changes slowly compared with the adjacent coherent transmission time blocks \cite{Liu_J_2022b}. Thus, we assume that the AoA, $\theta$, is fixed in the optimization. For the same reason, we also assume the Doppler shift, $f_d$, is a constant in the optimization \cite{berger2010signal}.

\begin{remark}
% We use CRB as the sensing metric rather than radar SINR or beampattern matching. This is because the fractional form in the radar SINR is non-convex and thus intractable. Additionally, both these two metrics concentrate on transmit side design, and neglect real target detection performances. In contrast, CRB focuses on the optimization of estimation performance, and implicitly constrains the beampattern direction.
We employ CRB as the sensing metric instead of radar SINR or beampattern matching. This choice is made due to the non-convex and thus intractable nature of the fractional form in radar SINR. Additionally, both of these metrics concentrate on transmit-side design, and neglect real target detection performance. In contrast, CRB prioritizes the optimization of estimation performance and implicitly constrains the beampattern direction.
\end{remark}

\subsection{Problem Formulation}
We aim to maximize the minimum EE among all users to guarantee fairness among users in this section. We formulate the max-min EE fairness problem and optimize the precoders $\mathbf{p}_c, \mathbf{p}_r, \{\mathbf{p}_k\}_{k \in \mathcal{K}}$, and the vector of common rate portions, $\mathbf{c}=\left[C_1, \cdots, C_K\right]^\top$. Therefore, the optimization problem is formulated as 
\begin{maxi!}|s|[2]                   % mini! = minimize 
    % {\mathbf{p}_c, \mathbf{p}_r, \{\mathbf{p}_k\}_{k \in \mathcal{K}}, \mathbf{c}}       
    {\substack{\mathbf{p}_c, \mathbf{p}_r, \mathbf{c}\\ \{\mathbf{p}_k\}_{k \in \mathcal{K}}} }  
    % optimization variable
    % {\frac{\min_{k \in \mathcal{K}} (C_k + R_k)}{{\sum_{k\in \mathcal{K}} \text{tr} (\mathbf{p}_k\mathbf{p}^H_k)} } 
    {\min_{k \in \mathcal{K}} \quad \frac{(C_k + R_k)}{{P_\mathsf{RF}+1/\kappa \,  (\norm{\boldsymbol{\Delta} \mathbf{p}_k}^2 + 1/K \norm{\boldsymbol{\Delta} \mathbf{p}_c}^2) }} \label{eq:fop1}}  % objective function and label
    {\label{eq:fp1}}             % label for opt problem
    % {\mathcal{P}1:} 
    % optimization result
    {} 
    \addConstraint{\norm{\boldsymbol{\Delta} \mathbf{p}_c}^2 + \sum_{k\in \mathcal{K}} \norm{\boldsymbol{\Delta} \mathbf{p}_k}^2 + \psi \norm{\boldsymbol{\Delta} \mathbf{p}_r}^2}{\leq P_\mathrm{t}, \label{eq:fp1c1}}
    \addConstraint{R_{\mathrm{c}, k} \geq \sum_{i \in \mathcal{K}} C_i, \quad \forall k \in \mathcal{K},}{\label{eq:fp1c2} }
    \addConstraint{C_k \geq 0, \quad \forall k \in \mathcal{K},}{\label{eq:fp1c3}}
    \addConstraint{R_k + C_k \geq R_\mathsf{th}, \quad \forall k \in \mathcal{K},}{ \label{eq:fp1c4}}
    \addConstraint{\mathsf{CRB}(\theta) \leq \rho,}{\label{eq:fp1c5}}%\addConstraint{\sum_{m=1}^M\left|P_{\mathsf{desired}}\left(\theta_m\right)-  \mathbf{a}^H\left(\theta_m\right) \mathbf{P} \mathbf{P}^H \mathbf{a}\left(\theta_m\right)\right|^2}{\leq \varepsilon, }
    % \addConstraint{\|\mathbf{w}\|^2_2}{ =1, \label{eq:op1c3}}
    % \addConstraint{\mathbf{w}^H \mathbf{H}_\mathrm{si} \mathbf{p}}{=0. \label{eq:op1c4}}
\end{maxi!}
where $R_\mathsf{th}$ is a parameter for the required rate for each user, $\rho$ is a parameter as the upper limit of CRB. Constraint \eqref{eq:fp1c1} limits the total transmit power of the ISAC satellite. \eqref{eq:fp1c2} guarantees the successful decoding of the common stream by all users. \eqref{eq:fp1c3} ensures all common rate portions are non-negative. \eqref{eq:fp1c4} is imposed to guarantee the QoS requirement of each user. To guarantee the target estimation performance, the CRB of the angle is upper limited as in \eqref{eq:fp1c5}.

\section{The Proposed Algorithm for Minimal Energy Efficiency Maximization}
In this section, we tackle the non-convexity of problem \eqref{eq:fp1}. To have computational efficiency in the problem solving, we transform the original problem to a SDP form. Since the non-convexity of the objective function \eqref{eq:fop1}  attributable to the fractional structure, we rewrite the objective function as a form of the difference of concave functions (DC). Subsequently, SCA is employed to transform the communication rate expression to a convex form. To handle the non-convexity in the fractional form of the objective function, we adopt Dinkelbach’s algorithm \cite{dinkelbach1967nonlinear} to transform the problem to an equivalent linear objective function. Subsequently, the objective function is transformed to a constraint. The constraints \eqref{eq:fp1c2} and \eqref{eq:fp1c4} are intractable due to the non-convex rate expressions, thus we introduce auxiliary variables and adopt SCA to have the linear approximation. The CRB constraint \eqref{eq:fp1c5} is transformed to a linear matrix inequality (LMI) by Schur Complement. To satisfy the rank-one constraint, a penalty function is added to the objective function. The details of the transformation and algorithm is given below.

% Specifically, the objective function \eqref{eq:fop1} and constraint \eqref{eq:fp1c5} are non-convex due to the fractional structure. \eqref{eq:fp1c2} and \eqref{eq:fp1c4} are intractable due to the non-convex rate expressions.

\bpara{SDP Transformation.} To make our algorithm computationally efficient, the original problem \eqref{eq:fp1} is first transformed into a SDP problem, given by
\begin{maxi!}|s|[2]                   % mini! = minimize  
    % { \mathbf{P}_c, \mathbf{P}_r, \{\mathbf{P}_k\}_{k \in \mathcal{K}}, \mathbf{c}}
    {\substack{\mathbf{P}_c, \mathbf{P}_r, \mathbf{c}, \\\{\mathbf{P}_k\}_{k \in \mathcal{K}} } }
    % optimization variable
    % {\frac{\min_{k \in \mathcal{K}} (C_k + R_k)}{{\sum_{k\in \mathcal{K}} \text{tr} (\mathbf{p}_k\mathbf{p}^H_k)} } 
    % {\min_{k \in \mathcal{K}} \quad \frac{(C_k + R_k (\mathbf{P}_c, \mathbf{P}_r, \mathbf{P}_k, \boldsymbol{\Delta}) \bigl)}{{P_\mathsf{RF}+1/\kappa \, \text{tr} (\boldsymbol{\Delta} \mathbf{P}_k \boldsymbol{\Delta} )}} \label{eq:sdpop}}  % objective function and label
    {{\min_{k \in \mathcal{K}} \quad \frac{(C_k + R_k ( \mathbf{P}_r, \mathbf{P}_k, \boldsymbol{\Delta}, \psi) \bigl)}{P_{\mathsf{chain}, k}(\mathbf{P}_k, \boldsymbol{\Delta})}} \label{eq:sdpop}}  % objective function and label
    {\label{eq:sdp}}             % label for opt problem
    % {\mathcal{P}1:} 
    % optimization result
    {} 
    \addConstraint{\operatorname{tr} ({\boldsymbol{\Delta} \mathbf{P} \mathbf{P}^H \boldsymbol{\Delta}} ) }{\leq P_\mathrm{t}, \label{eq:sdpc1}}
    \addConstraint{\mathbf{P}_c \succeq 0, \mathbf{P}_r \succeq 0, \mathbf{P}_k \succeq 0, \forall k \in \mathcal{K}  }{ \label{eq:sdpc2}}
    % \addConstraint{
    % \operatorname{rank}(\mathbf{P}_c) = 1, \operatorname{rank}(\mathbf{P}_k) = 1, \operatorname{rank}(\mathbf{P}_k) = 1, \forall k \in \mathcal{K}}{ \label{eq:sdpc3}}
    \addConstraint{
    \operatorname{rank}(\mathbf{P}_c) = 1, \operatorname{rank}(\mathbf{P}_r) = 1, \operatorname{rank}(\mathbf{P}_k) = 1,} \nonumber
    {}
    \addConstraint{}{\forall k \in \mathcal{K}, \label{eq:sdpc3}}
    \addConstraint{\log_2\left(1+ \Gamma_{c,k} \right) \geq \sum_{i \in \mathcal{K} }C_i, \quad \forall k \in \mathcal{K}, }{\label{eq:sdpc4} }
    \addConstraint{C_k \geq 0, \quad \forall k \in \mathcal{K},}{\label{eq:sdpc5}}
    \addConstraint{\log_2\left(1+ \Gamma_{k} \right) \geq r_k, \quad \forall k \in \mathcal{K},}{ \label{eq:sdpc6}}
    \addConstraint{C_k + r_k \geq R_\mathsf{th} \quad \forall k \in \mathcal{K},}{\label{eq:sdpc7}}
    \addConstraint{\mathsf{CRB}(\theta) \leq \rho,}{\label{eq:sdpc8}}%\addConstraint{\sum_{m=1}^M\left|P_{\mathsf{desired}}\left(\theta_m\right)-  \mathbf{a}^H\left(\theta_m\right) \mathbf{P} \mathbf{P}^H \mathbf{a}\left(\theta_m\right)\right|^2}{\leq \varepsilon, }
    % \addConstraint{\|\mathbf{w}\|^2_2}{ =1, \label{eq:op1c3}}
    % \addConstraint{\mathbf{w}^H \mathbf{H}_\mathrm{si} \mathbf{P}}{=0. \label{eq:op1c4}}
\end{maxi!}
where we define $\mathbf{P}_c = \mathbf{p}_c \mathbf{p}_c^H, \mathbf{P}_r = \mathbf{p}_r \mathbf{p}_r^H, \mathbf{P}_k = \mathbf{p}_k \mathbf{p}_k^H, \mathbf{H}_k = \mathbf{h}_k \mathbf{h}_k^H$ for the SDP transformation; $\mathbf{r} = [r_1, \cdots, r_K]^\top$ is an auxiliary variable; {$P_{\mathsf{chain}, k}(\mathbf{P}_k, \boldsymbol{\Delta}) = {P_\mathsf{RF}+1/\kappa \, (\text{tr} (\boldsymbol{\Delta} \mathbf{P}_k \boldsymbol{\Delta} ) + 1/K \left(\boldsymbol{\Delta} \mathbf{P}_c \boldsymbol{\Delta} )    \right)}$, }
\begin{subequations}
\begin{equation}
    \Gamma_{c,k} \! = \! \frac{\operatorname{tr}\left(\mathbf{H}_k \boldsymbol{\Delta} \mathbf{P}_c \boldsymbol{\Delta} \right)}{\sum\limits_{j \in \mathcal{K}} \operatorname{tr}\left(\mathbf{H}_k \boldsymbol{\Delta} \mathbf{P}_j \boldsymbol{\Delta} \right) \! + \! \psi \operatorname{tr}\left(\mathbf{H}_k \boldsymbol{\Delta} \mathbf{P}_r \boldsymbol{\Delta} \right) \! + \! \operatorname{tr}\left(\mathbf{H}_k \boldsymbol{\Sigma}  \right) \! + \! \sigma_k^2},
\end{equation}
\begin{equation}
    \Gamma_{k}\! =\! \frac{\operatorname{tr}\left(\mathbf{H}_k \boldsymbol{\Delta} \mathbf{P}_k \boldsymbol{\Delta}\right)}{\sum\limits_{\substack{j \in \mathcal{K}, \\ j \neq k}} \operatorname{tr}\left(\mathbf{H}_k \boldsymbol{\Delta} \mathbf{P}_j \boldsymbol{\Delta} \right)\!+ \! \psi \operatorname{tr}\left(\mathbf{H}_k \boldsymbol{\Delta} \mathbf{P}_r \boldsymbol{\Delta} \right) \! + \! \operatorname{tr}\left(\mathbf{H}_k \boldsymbol{\Sigma}  \right) \! + \! \sigma_k^2},
\end{equation}
\end{subequations}
and $R_k(\mathbf{P}_r, \mathbf{P}_k, \boldsymbol{\Delta}, \psi)$ is given in \eqref{eq:rksdp}.
\begin{figure*}[tb]
\begin{equation}
\label{eq:rksdp}
\begin{aligned}
{
    R_k(\mathbf{P}_r, \mathbf{P}_k, \boldsymbol{\Delta}, \psi) }
    & {= \log_2 \bigl(  {\sum\limits_{j \in \mathcal{K}} \operatorname{tr}\left(\mathbf{H}_k \boldsymbol{\Delta} \mathbf{P}_j \boldsymbol{\Delta} \right) \! + \!  \psi \operatorname{tr}\left(\mathbf{H}_k \boldsymbol{\Delta} \mathbf{P}_r \boldsymbol{\Delta} \right) \! + \! \operatorname{tr}\left(\mathbf{H}_k \boldsymbol{\Sigma}  \right) \! + \! \sigma_k^2}  \bigl) }\\ 
    & {\quad - \log_2 \bigl( {\sum\limits_{\substack{j \in \mathcal{K}, j \neq k}} \operatorname{tr}\left(\mathbf{H}_k \boldsymbol{\Delta} \mathbf{P}_j \boldsymbol{\Delta} \right) \! + \! \psi \operatorname{tr}\left(\mathbf{H}_k \boldsymbol{\Delta} \mathbf{P}_r \boldsymbol{\Delta} \right) \! + \! \operatorname{tr}\left(\mathbf{H}_k \boldsymbol{\Sigma}  \right) \! + \! \sigma_k^2}  \bigl)}
    \end{aligned}
\end{equation}
\hrulefill
\end{figure*}

\bpara{Objective Function Transformation by SCA and Dinkelbach's Method.}
Since the rate expression in the numerator, $R_{\mathsf{total},k}(\mathbf{P}_r, \mathbf{P}_k, \boldsymbol{\Delta}, C_k, \psi)$, of the objective function \eqref{eq:sdp} is intractable, and it is a classic DC programming problem, we can tackle this problem by SCA method \cite{yang2018energy}. Thus, the numerator in \eqref{eq:sdpop} is rewritten as
\begin{equation}
\begin{aligned}
    R_{\mathsf{total},k}(& \mathbf{P}_r, \mathbf{P}_k, \boldsymbol{\Delta}, C_k, \psi) 
     = C_k + \log_2 (1 + \Gamma_k)\\
    & = f_k( \mathbf{P}_r, \mathbf{P}_k, \boldsymbol{\Delta}, C_k, \psi) - g_k(\mathbf{P}_r, \mathbf{P}_k, \boldsymbol{\Delta}, \psi),
\end{aligned}
\end{equation}
where
\begin{subequations}
\begin{equation}
\begin{aligned}
    f_k( \mathbf{P}_r, \mathbf{P}_k, & \boldsymbol{\Delta}, C_k, \psi)  \triangleq  C_k \!  +  \log_2 \bigl( \sum_{j \in \mathcal{K}} \operatorname{tr}\left(\mathbf{H}_k \boldsymbol{\Delta} \mathbf{P}_j \boldsymbol{\Delta} \right) \\
    & + \! \psi \operatorname{tr}\left(\mathbf{H}_k \boldsymbol{\Delta} \mathbf{P}_r \boldsymbol{\Delta} \right) \! + \! \operatorname{tr}\left(\mathbf{H}_k \boldsymbol{\Sigma}  \right) \! + \! \sigma_k^2  \bigl),
\end{aligned}
\end{equation}
% \begin{equation}
% \begin{aligned}
%     g_k(\mathbf{P}_r, \mathbf{P}_k, \boldsymbol{\Delta}, \psi) & \! = \! \log_2 \bigl( {\sum_{\substack{j \in \mathcal{K},  j \neq k}} \operatorname{tr}(\mathbf{H}_k \boldsymbol{\Delta} \mathbf{P}_j \boldsymbol{\Delta} ) \! + \! \psi \operatorname{tr}(\mathbf{H}_k \boldsymbol{\Delta} \mathbf{P}_r \boldsymbol{\Delta} )} \! \\ 
%     & \quad + \! \operatorname{tr}\left(\mathbf{H}_k \boldsymbol{\Sigma}  \right) \! + \! \sigma_k^2  \bigl).
% \end{aligned}
% \end{equation}
\begin{equation}
\begin{aligned}
    g_k( \mathbf{P}_r, \mathbf{P}_k, & \boldsymbol{\Delta},  \psi) \! \triangleq \! \log_2 \bigl( \sum_{\substack{j \in \mathcal{K},  j \neq k}} \operatorname{tr}(\mathbf{H}_k \boldsymbol{\Delta} \mathbf{P}_j \boldsymbol{\Delta} ) \! \\
    & + \! \psi \operatorname{tr}(\mathbf{H}_k \boldsymbol{\Delta} \mathbf{P}_r \boldsymbol{\Delta} ) \!  + \! \operatorname{tr}\left(\mathbf{H}_k \boldsymbol{\Sigma}  \right) \! + \! \sigma_k^2  \bigl).
\end{aligned}
\end{equation}
\end{subequations}
By using the SCA, the concave function $g_k(\mathbf{P}_r, \mathbf{P}_k, \boldsymbol{\Delta}, \psi)$ is upper bounded by its linear approximation at any feasible point with respect to $\mathbf{P}_r$ and $\mathbf{P}_k$, rewritten as
\begin{equation}
\begin{aligned}
     & g_k( \mathbf{P}_r, \mathbf{P}_k, \boldsymbol{\Delta}, \psi)  \leq g_k(\mathbf{P}_r^{(t)}, \mathbf{P}_k^{(t)}, \boldsymbol{\Delta}, \psi) \!  \\ & +  \!  \sum\limits_{\substack{j \in \mathcal{K}, j \neq k}} \! \operatorname{tr} (\grad^H_{\mathbf{P}_j} g_k(\mathbf{P}_r^{(t)}, \mathbf{P}_k^{(t)}, \boldsymbol{\Delta}, \psi)  (\mathbf{P}_j - \mathbf{P}_j^{(t)}) ) 
    \\ & + \operatorname{tr} (\grad^H_{\mathbf{P}_r} g_k(\mathbf{P}_r^{(t)}, \mathbf{P}_k^{(t)}, \boldsymbol{\Delta}, \psi)   (\mathbf{P}_r - \mathbf{P}_r^{(t)}) \\
    & \triangleq \widetilde{g}(\mathbf{P}_r^{(t)}, \mathbf{P}_k^{(t)}, \boldsymbol{\Delta}, \psi),
\end{aligned}
\end{equation}
where $t$ represents the $t^\text{th}$ SCA iteration, and
% \begin{equation}
%     \grad^H_{\mathbf{P}_j} g_k(\mathbf{P}_r^{(t)}, \mathbf{P}_k^{(t)}, \boldsymbol{\Delta}, \psi) \! = \!\grad^H_{\mathbf{P}_r} g_k(\mathbf{P}_r^{(t)}, \mathbf{P}_k^{(t)}, \boldsymbol{\Delta}, \psi) \! = \!\frac{\boldsymbol{\Delta} \mathbf{H}_k \boldsymbol{\Delta}}{\ln2},\! \forall j.
% \end{equation}
\begin{align}
    \grad^H_{\mathbf{P}_j} g_k(\mathbf{P}_r^{(t)}, \mathbf{P}_k^{(t)}, \boldsymbol{\Delta}, \psi)  
    & = \grad^H_{\mathbf{P}_r} g_k(\mathbf{P}_r^{(t)}, \mathbf{P}_k^{(t)}, \boldsymbol{\Delta}, \psi) \\ \nonumber
    & = \frac{\boldsymbol{\Delta} \mathbf{H}_k \boldsymbol{\Delta}}{\ln2}, \quad \forall j.
\end{align}
Since $\widetilde{g}(\mathbf{P}_r^{(t)}, \mathbf{P}_k^{(t)}, \boldsymbol{\Delta}, \psi)$ is now an affine function, $\widetilde{R}_{\mathsf{total},k}(\mathbf{P}_r^{(t)}, \mathbf{P}_k^{(t)}, \boldsymbol{\Delta}, C_k^{(t)}, \psi) \triangleq f_k( \mathbf{P}_r, \mathbf{P}_k, \boldsymbol{\Delta}, C_k, \psi) - \widetilde{g}(\mathbf{P}_r^{(t)}, \mathbf{P}_k^{(t)}, \boldsymbol{\Delta}, \psi)$ is a concave function. Therefore, the transformed problem is as follows
\begin{maxi!}|s|[2]                   % mini! = minimize  
    % { \mathbf{P}_c, \mathbf{P}_r, \{\mathbf{P}_k\}_{k \in \mathcal{K}}, \mathbf{c}}
    {\substack{\mathbf{P}_c, \mathbf{P}_r, \mathbf{c}, \\\{\mathbf{P}_k\}_{k \in \mathcal{K}} } }
    % optimization variable
    % {\frac{\min_{k \in \mathcal{K}} (C_k + R_k)}{{\sum_{k\in \mathcal{K}} \text{tr} (\mathbf{p}_k\mathbf{p}^H_k)} } 
    {\min_{k \in \mathcal{K}} \quad \frac{\widetilde{R}_{\mathsf{total},k}(\mathbf{P}_r, \mathbf{P}_k, \boldsymbol{\Delta}, C_k, \psi)}{P_{\mathsf{chain}, k}(\mathbf{P}_k, \boldsymbol{\Delta})} \label{eq:dcop}}  % objective function and label
    {\label{eq:dc}}             % label for opt problem
    % {\mathcal{P}1:} 
    % optimization result
    {} 
    \addConstraint{\eqref{eq:sdpc1}-\eqref{eq:sdpc8}.}{} \nonumber
\end{maxi!}

Subsequently, this fractional programming (FP) problem \eqref{eq:dc} can be iteratively solved by the Dinkelbach's algorithm \cite{dinkelbach1967nonlinear}, the transformed problem with integral expression in the objective function \eqref{eq:dcop} is rewritten as
\begin{maxi!}|s|[2]                   % mini! = minimize  
    % { \mathbf{P}_c, \mathbf{P}_r, \{\mathbf{P}_k\}_{k \in \mathcal{K}}, \mathbf{c}}
    {\substack{\mathbf{P}_c, \mathbf{P}_r, \mathbf{c}, \\\{\mathbf{P}_k\}_{k \in \mathcal{K}} } }
    % optimization variable
    % {\frac{\min_{k \in \mathcal{K}} (C_k + R_k)}{{\sum_{k\in \mathcal{K}} \text{tr} (\mathbf{p}_k\mathbf{p}^H_k)} } 
    {\!\min_{k \in \mathcal{K}}  {\widetilde{R}_{\mathsf{total},k}(\mathbf{P}_r, \mathbf{P}_k, \boldsymbol{\Delta}, C_k, \psi)} 
    \! - \! \lambda \left( P_{\mathsf{chain}, k}(\mathbf{P}_k, \boldsymbol{\Delta}) \right) \label{eq:dinkop}}  % objective function and label
    {\label{eq:dink}}             % label for opt problem
    % {\mathcal{P}1:} 
    % optimization result
    {} 
    \addConstraint{\eqref{eq:sdpc1}-\eqref{eq:sdpc8}.}{} \nonumber
\end{maxi!}
Next, the variable $\lambda$ is updated as follow
\begin{equation}
    \lambda^{(t+1)} = \min_{k \in \mathcal{K}} \quad \frac{\widetilde{R}_{\mathsf{total},k}(\mathbf{P}_r^{(t)}, \mathbf{P}_k^{(t)}, \boldsymbol{\Delta}, C_k^{(t)}, \psi)}{P_{\mathsf{chain}, k}(\mathbf{P}_k^{(t)}, \boldsymbol{\Delta})},
\end{equation}
where $t$ denotes the $t^{\text{th}}$ iteration in the optimization process.

To make max-min objective function \eqref{eq:dinkop} smooth, we introduce an auxiliary variable $\mu$ \cite{zappone2017globally}, and the equivalent problem is as follows
\begin{maxi!}|s|[2]                   % mini! = minimize  
    % { \mathbf{P}_c, \mathbf{P}_r, \{\mathbf{P}_k\}_{k \in \mathcal{K}}, \mathbf{c}}
    {\substack{\mathbf{P}_c, \mathbf{P}_r, \mathbf{c}, \\\{\mathbf{P}_k\}_{k \in \mathcal{K}}, \mu } }
    % optimization variable
    % {\frac{\min_{k \in \mathcal{K}} (C_k + R_k)}{{\sum_{k\in \mathcal{K}} \text{tr} (\mathbf{p}_k\mathbf{p}^H_k)} } 
    {\mu  \label{eq:minop}}  % objective function and label
    {\label{eq:minp}}             % label for opt problem
    % {\mathcal{P}1:} 
    % optimization result
    {} 
    \addConstraint{{\widetilde{R}_{\mathsf{total},k}(\mathbf{P}_r, \mathbf{P}_k, \boldsymbol{\Delta}, C_k, \psi)} 
    - \lambda \left( P_{\mathsf{chain}, k}(\mathbf{P}_k, \boldsymbol{\Delta}) \right)  }{} \nonumber
    \addConstraint{\geq \mu, \forall k \in \mathcal{K}}{ \label{eq:minc1}}
    \addConstraint{\eqref{eq:sdpc1}-\eqref{eq:sdpc8}.}{} \nonumber
\end{maxi!}
\textbf{Rate Constraints Transformation by SCA.} To tackle the non-convexity in \eqref{eq:sdpc4} and \eqref{eq:sdpc6}, we utilize the SCA technique to transform these two constraints. Inspired by \cite{yin2022ratewcnc}, we first introduce slack variables $\eta_{c,k}, \eta_{k}, \beta_{c,k}, \beta_{k}, \forall k \in \mathcal{K}$. Thus, \eqref{eq:sdpc4} is transformed to
\begin{align}
& \label{eq:etaminusbetack}\eta_{c, k}-\beta_{c, k} \geq \sum_{i \in \mathcal{K}} C_i \log 2,\\
& \label{eq:eetack}e^{\eta_{c, k}} \leq \operatorname{tr}\left(\mathbf{H}_k \boldsymbol{\Delta} \mathbf{P}_c \boldsymbol{\Delta} \right)\! + \! \psi \operatorname{tr}\left(\mathbf{H}_k \boldsymbol{\Delta} \mathbf{P}_r \boldsymbol{\Delta} \right) \! + \! \operatorname{tr}\left(\mathbf{H}_k \boldsymbol{\Sigma}  \right) \\ \nonumber
& \qquad \quad \!+ \! \sum\limits_{j \in \mathcal{K}} \operatorname{tr}\left(\mathbf{H}_k \boldsymbol{\Delta} \mathbf{P}_j \boldsymbol{\Delta} \right)  \! + \! \sigma_k^2,  \\
& \label{eq:ebetack} e^{\beta_{c, k}} \! \geq \!
\sum\limits_{j \in \mathcal{K}} \operatorname{tr}\left(\mathbf{H}_k \boldsymbol{\Delta} \mathbf{P}_j \boldsymbol{\Delta} \right) \! + \! \psi \operatorname{tr}\left(\mathbf{H}_k \boldsymbol{\Delta} \mathbf{P}_r \boldsymbol{\Delta} \right) \\ \nonumber
&\qquad \quad  \! + \! \operatorname{tr}\left(\mathbf{H}_k \boldsymbol{\Sigma}  \right) \! + \! \sigma_k^2,
\end{align}
% \vspace{-5pt}
where $\forall k \in \mathcal{K}$. Next, \eqref{eq:sdpc6} is transformed to 
\vspace{-8pt}
\begin{align}
& \label{eq:etaminusbetak} \eta_{k}-\beta_{k} \geq r_k \log 2,\\
& e^{\eta_{k}} \leq 
\sum\limits_{j \in \mathcal{K}} \operatorname{tr}\left(\mathbf{H}_k \boldsymbol{\Delta} \mathbf{P}_j \boldsymbol{\Delta} \right) \! + \! \psi \operatorname{tr}\left(\mathbf{H}_k \boldsymbol{\Delta} \mathbf{P}_r \boldsymbol{\Delta} \right) \! + \! \operatorname{tr}\left(\mathbf{H}_k \boldsymbol{\Sigma}  \right) \! + \! \sigma_k^2, \label{eq:eetak} \\
& e^{\beta_{k}} \! \geq \!
\sum\limits_{\substack{j \in \mathcal{K},\\ j \neq k}} \operatorname{tr}\left(\mathbf{H}_k \boldsymbol{\Delta} \mathbf{P}_j \boldsymbol{\Delta} \right) \! + \! \psi \operatorname{tr}\left(\mathbf{H}_k \boldsymbol{\Delta} \mathbf{P}_r \boldsymbol{\Delta} \right) \! + \! \operatorname{tr}\left(\mathbf{H}_k \boldsymbol{\Sigma}  \right) \! + \! \sigma_k^2,
\label{eq:ebetak}
\end{align}
% \vspace{-5pt}
% \begin{equation}
% \label{eq:etaminusbetak} 
% \eta_{k}-\beta_{k} \geq r_k \log 2,
% \end{equation}
% \begin{equation}
% \begin{aligned}
% e^{\eta_{k}} \leq 
% \sum\limits_{j \in \mathcal{K}} \operatorname{tr}\left(\mathbf{H}_k \boldsymbol{\Delta} \mathbf{P}_j \boldsymbol{\Delta} \right) \! + \! \psi \operatorname{tr}\left(\mathbf{H}_k \boldsymbol{\Delta} \mathbf{P}_r \boldsymbol{\Delta} \right) \! + \! \operatorname{tr}\left(\mathbf{H}_k \boldsymbol{\Sigma}  \right) \! + \! \sigma_k^2, \label{eq:eetak}
% \end{aligned}
% \end{equation}
% \begin{equation}
% \begin{aligned}
% e^{\beta_{k}} \! \geq \!
% \sum\limits_{\substack{j \in \mathcal{K},\\ j \neq k}} \operatorname{tr}\left(\mathbf{H}_k \boldsymbol{\Delta} \mathbf{P}_j \boldsymbol{\Delta} \right) \! + \! \psi \operatorname{tr}\left(\mathbf{H}_k \boldsymbol{\Delta} \mathbf{P}_r \boldsymbol{\Delta} \right) \! + \! \operatorname{tr}\left(\mathbf{H}_k \boldsymbol{\Sigma}  \right) \! + \! \sigma_k^2,
% \label{eq:ebetak}
% \end{aligned}
% \end{equation}
where $\forall k \in \mathcal{K}$. However, \eqref{eq:ebetack} and \eqref{eq:ebetak} are still non-convex, thus first-order Taylor approximation is utilized to tackle these two constraints, given by
\begin{equation}
\begin{aligned}
\label{eq:ebb1ck}
& \sum\limits_{j \in \mathcal{K}} \operatorname{tr}\left(\mathbf{H}_k \boldsymbol{\Delta} \mathbf{P}_j \boldsymbol{\Delta} \right) \! + \! \psi \operatorname{tr}\left(\mathbf{H}_k \boldsymbol{\Delta} \mathbf{P}_r \boldsymbol{\Delta} \right) \! + \! \operatorname{tr}\left(\mathbf{H}_k \boldsymbol{\Sigma}  \right)  \\
& \qquad \! + \! \sigma_k^2 \leq e^{\beta_{c, k}^{(t)}}\left(\beta_{c, k}-\beta_{c, k}^{(t)}+1\right),
\end{aligned}
\end{equation}
\begin{equation}
\begin{aligned}
& \label{eq:ebb1k}\sum\limits_{\substack{j \in \mathcal{K}, j \neq k}} \operatorname{tr}\left(\mathbf{H}_k \boldsymbol{\Delta} \mathbf{P}_j \boldsymbol{\Delta} \right) \! + \! \psi \operatorname{tr}\left(\mathbf{H}_k \boldsymbol{\Delta} \mathbf{P}_r \boldsymbol{\Delta} \right)   \\
& \qquad \! + \! \operatorname{tr}\left(\mathbf{H}_k \boldsymbol{\Sigma}  \right) \! + \! \sigma_k^2 \leq e^{\beta_k^{(t)}}\left(\beta_k-\beta_k^{(t)}+1\right).
\end{aligned}
\end{equation}
%\begin{align}
%& \label{eq:ebb1ck}\sum\limits_{j \in \mathcal{K}} \operatorname{tr}\left(\mathbf{H}_k \boldsymbol{\Delta} \mathbf{P}_j \boldsymbol{\Delta} \right) \! + \! \psi \operatorname{tr}\left(\mathbf{H}_k \boldsymbol{\Delta} \mathbf{P}_r \boldsymbol{\Delta} \right) \! + \! \operatorname{tr}\left(\mathbf{H}_k \boldsymbol{\Sigma}  \right) \! + \! \sigma_k^2 \\ \nonumber
%& \leq e^{\beta_{c, k}^{(t)}}\left(\beta_{c, k}-\beta_{c, k}^{(t)}+1\right), \\
%&  \label{eq:ebb1k}\sum\limits_{\substack{j \in \mathcal{K}, j \neq k}} \operatorname{tr}\left(\mathbf{H}_k \boldsymbol{\Delta} \mathbf{P}_j \boldsymbol{\Delta} \right) \! + \! \psi \operatorname{tr}\left(\mathbf{H}_k \boldsymbol{\Delta} \mathbf{P}_r \boldsymbol{\Delta} \right) \! + \! \operatorname{tr}\left(\mathbf{H}_k \boldsymbol{\Sigma}  \right) \! + \! \sigma_k^2 \\ \nonumber
%& \leq e^{\beta_k^{(t)}}\left(\beta_k-\beta_k^{(t)}+1\right).
%\end{align}
\eqref{eq:eetack} and \eqref{eq:eetak} belong to generalized nonlinear convex programming, and result in high computational complexity. To efficiently cope with \eqref{eq:eetack} and \eqref{eq:eetak}, we introduce auxiliary variables $\tau_{c, k}, \tau_{k}$, and then compute first-order Taylor approximation. Therefore, \eqref{eq:eetack} and \eqref{eq:eetak} are rewritten as 
\begin{align}
& \label{eq:tauck} \tau_{c, k} \leq \operatorname{tr}\left(\mathbf{H}_k \boldsymbol{\Delta} \mathbf{P}_c \boldsymbol{\Delta} \right) + \! \psi \operatorname{tr}\left(\mathbf{H}_k \boldsymbol{\Delta} \mathbf{P}_r \boldsymbol{\Delta} \right) \! + \! \operatorname{tr}\left(\mathbf{H}_k \boldsymbol{\Sigma}  \right)\\ \nonumber
& \qquad \quad  \! + \!\sum\limits_{j \in \mathcal{K}} \operatorname{tr}\left(\mathbf{H}_k \boldsymbol{\Delta} \mathbf{P}_j \boldsymbol{\Delta} \right) \! \! + \! \sigma_k^2, \\
% & \tau_{c, k} \log \left(\tau_{c, k}\right) \geq \tau_{c, k} \eta_{c, k}, \\
& \label{eq:taucketack}\tau_{c, k}^{(t)} \log \left(\tau_{c, k}^{(t)}\right) \! + \! \left(\tau_{c, k} \! - \! \tau_{c, k}^{(t)}\right)\left[\log \left(\tau_{c, k}^{(t)}\right) \! + \! 1\right] \! \geq \! \tau_{c, k} \eta_{c, k}, \\
& \label{eq:tauk} \tau_k \! \leq \! \sum\limits_{j \in \mathcal{K}} \operatorname{tr}\left(\mathbf{H}_k \boldsymbol{\Delta} \mathbf{P}_j \boldsymbol{\Delta} \right) \! + \! \psi \operatorname{tr}\left(\mathbf{H}_k \boldsymbol{\Delta} \mathbf{P}_r \boldsymbol{\Delta} \right) \! + \! \operatorname{tr}\left(\mathbf{H}_k \boldsymbol{\Sigma}  \right) \! + \! \sigma_k^2, \\
% & \tau_k \log \left(\tau_k\right) \geq \tau_k \eta_k.
& \label{eq:tauketak}\tau_k^{(t)} \log \left(\tau_k^{(t)}\right)+\left(\tau_k-\tau_k^{(t)}\right)\left[\log \left(\tau_k^{(t)}\right)+1\right] \geq \tau_k \eta_k.
\end{align}
{
To increase efficiency and numerical robustness of the solution, we transform \eqref{eq:taucketack} and \eqref{eq:tauketak} to second-order cone (SOC) forms. Taking the transformation of \eqref{eq:taucketack} as an example, \eqref{eq:taucketack} can be rewritten by multiplying $4$ in both sides as 
\begin{equation}
    \label{eq:taucketack1}
    -4 \tau_{c, k} (\eta_{c, k} - ( \log \left(\tau_{c, k}^{(t)}\right) \! + \! 1 ) ) \geq 
    4 \tau_{c, k}^{(t)}.   
    % 4 \tau_{c, k} \left[\log \left(\tau_{c, k}^{(t)}\right) \! + \! 1\right] - 4 \tau_{c, k} \eta_{c, k} \geq 
    % 4 \tau_{c, k}^{(t)},   
    % \tau_{c, k}^{(t)} \log \left(\tau_{c, k}^{(t)}\right) \! + \! \left(\tau_{c, k} \! - \! \tau_{c, k}^{(t)}\right)\left[\log \left(\tau_{c, k}^{(t)}\right) \! + \! 1\right] \! \geq \! \tau_{c, k} \eta_{c, k},
\end{equation}
Through completing the square in both sides of the inequality \eqref{eq:taucketack1}, it is further transformed as 
\begin{equation}
    \label{eq:taucketack2}
    \begin{aligned}
    & \left( \tau_{c, k} +  \left(\eta_{c, k} - ( \log \left(\tau_{c, k}^{(t)}\right) \! + \! 1 ) \right) \right)^2 + 4 \tau_{c, k}^{(t)} \\
    & \leq \left( \tau_{c, k} - \left(\eta_{c, k} - ( \log \left(\tau_{c, k}^{(t)}\right) \! + \! 1 ) \right) \right)^2.
    \end{aligned}
\end{equation}}
Therefore, the equivalent SOC forms of \eqref{eq:taucketack} and \eqref{eq:tauketak} are as follows
\begin{align}
& \label{eq:socck} \bigl\| \bigl[\tau_{c, k}+\eta_{c, k}-\left(\log \left(\tau_{c, k}^{(t)}\right)+1\right), 2 \sqrt{\tau_{c, k}^{(t)}} \bigl] \bigl\|_2 \\ \nonumber
& \leq \tau_{c, k}-\eta_{c, k}+\log \left(\tau_{c, k}^{(t)}\right)+1,\\
& \label{eq:sock} \bigl\|\bigl[ \tau_k+\eta_k-\left(\log \left(\tau_k^{(t)}\right)+1\right), 2 \sqrt{\tau_k^{(t)}} \bigl] \bigl\|_2 \\ \nonumber
& \leq \tau_k-\eta_k+\log \left(\tau_k^{(t)}\right)+1.
\end{align}
Thus, the transformed problem is rewritten as
\begin{maxi!}|s|[2]                   % mini! = minimize  
    {\substack{\mathbf{P}_c, \mathbf{P}_r, \mathbf{c}, \\\{\mathbf{P}_k\}_{k \in \mathcal{K}}, \mu } }
    % optimization variable
    {\mu  \label{eq:op2}}  % objective function and label
    {\label{eq:p2}}             % label for opt problem
    % {\mathcal{P}1:} 
    % optimization result
    {} 
    \addConstraint{\eqref{eq:sdpc1}-\eqref{eq:sdpc3}, \eqref{eq:sdpc5}, \eqref{eq:sdpc7}, \eqref{eq:sdpc8}, \eqref{eq:minc1},  
    }{} \nonumber
    \addConstraint{\eqref{eq:etaminusbetack}, \eqref{eq:etaminusbetak},  \eqref{eq:ebb1ck}, \eqref{eq:ebb1k}, \eqref{eq:tauck}, \eqref{eq:tauk}, \eqref{eq:socck}, \eqref{eq:sock}.
    }{} \nonumber
\end{maxi!}
%\bpara{CRB Transformation by Schur Complement.} 
\textbf{CRB Transformation by Schur Complement.} To handle the non-convexity in the $\mathsf{CRB}(\theta)$ \eqref{eq:crb} due to the fractional structure, we first reformulate it as 
\begin{equation}
\operatorname{tr}\left(\dot{\mathbf{A}}(\theta) \mathbf{R}_{\mathsf{q}} \dot{\mathbf{A}}^H(\theta)\right) \! - \! \frac{\left|\operatorname{tr}\left(\mathbf{A}(\theta) \mathbf{R}_{\mathsf{q}} \dot{\mathbf{A}}^H(\theta)\right)\right|^2}{\operatorname{tr}\left({\mathbf{A}}(\theta) \mathbf{R}_{\mathsf{q}} {\mathbf{A}}^H(\theta)\right)} \! - \! \frac{1}{2 \rho \mathsf{SNR}} \! \geq \! 0.
\end{equation}
By leveraging Schur complement \cite{zhang2006schur,Liu_J_2022b}, we have the following linear matrix inequality, given by
\begin{equation}
\label{eq:crbschur}
\left[\begin{array}{ll}
\operatorname{tr}\left( \dot{\mathbf{A}} \mathbf{R}_\mathsf{q} \dot{\mathbf{A}}^H \right)- \frac{1}{2 \rho \mathsf{SNR}} & \operatorname{tr}\left( \mathbf{A} \mathbf{R}_\mathsf{q} \dot{\mathbf{A}}^H\right) \\
\operatorname{tr}\left(\dot{\mathbf{A}}   \mathbf{R}_\mathsf{q} \mathbf{A}^H \right) & \operatorname{tr}\left( \mathbf{A} \mathbf{R}_\mathsf{q} \mathbf{A}^H\right)
\end{array}\right] \succeq \mathbf{0}.
\end{equation}
Therefore, the reformulated problem is written as
\begin{maxi!}|s|[2]                   % mini! = minimize  
    {\substack{\mathbf{P}_c, \mathbf{P}_r, \mathbf{c}, \\\{\mathbf{P}_k\}_{k \in \mathcal{K}}, \mu } }
    % optimization variable
    {\mu  \label{eq:op3}}  % objective function and label
    {\label{eq:p3}}             % label for opt problem
    % {\mathcal{P}1:} 
    % optimization result
    {} 
    \addConstraint{\eqref{eq:sdpc1}-\eqref{eq:sdpc3}, \eqref{eq:sdpc5}, \eqref{eq:sdpc7}, \eqref{eq:minc1},  \eqref{eq:etaminusbetack},
    }{} \nonumber
    \addConstraint{ \eqref{eq:etaminusbetak},  \eqref{eq:ebb1ck}, \eqref{eq:ebb1k}, \eqref{eq:tauck}, \eqref{eq:tauk}, \eqref{eq:socck}, \eqref{eq:sock}, \eqref{eq:crbschur}.
    }{} \nonumber
\end{maxi!}

\bpara{Rank-one Constraint Transformation by Penalty Function.}
The rank-one constraints \eqref{eq:sdpc3} are non-convex, we adopt the penalty function and a corresponding iterative method to tackle this issue \cite{yin2022rate, boyd2004convex}. Given that a rank-one matrix implies the presence of only a single nonzero eigenvalue, we can rewrite the non-convex constraints \eqref{eq:sdpc3} in an alternative form as
\begin{subequations}
    \begin{align}
    \operatorname{tr}(\mathbf{P}_c) - \chi_\mathsf{max}(\mathbf{P}_c) & = 0,\\
    \operatorname{tr}(\mathbf{P}_r) - \chi_\mathsf{max}(\mathbf{P}_r) & = 0, \\
    \operatorname{tr}(\mathbf{P}_k) - \chi_\mathsf{max}(\mathbf{P}_k) = 0,  & \quad \forall k \in \mathcal{K},
    \end{align}
\end{subequations}
where $\chi_\mathsf{max}()$ denotes the maximum eigenvalue of a matrix. Subsequently, we formulate a penalty function to incorporate these constraints into the objective function \eqref{eq:op3}, as follows
\begin{maxi!}|s|[2]                   % mini! = minimize  
    {\substack{\mathbf{P}_c, \mathbf{P}_r, \mathbf{c}, \\\{\mathbf{P}_k\}_{k \in \mathcal{K}}, \mu } }
    % optimization variable
    {\mu \!-\! \xi \bigl( [\operatorname{tr}(\mathbf{P}_c)\! - \!\chi_\mathsf{max}(\mathbf{P}_c)] \!+\! \psi [\operatorname{tr}(\mathbf{P}_r) \! - \!\chi_\mathsf{max}(\mathbf{P}_r) ]
    \label{eq:op4}}  % objective function and label
    {\label{eq:p4}}             % label for opt problem
    % {\mathcal{P}1:} 
    % optimization result
    {} \nonumber
    \breakObjective{+ \sum_{k \in \mathcal{K}} [\operatorname{tr}(\mathbf{P}_k) - \chi_\mathsf{max}(\mathbf{P}_k)] \bigl)}
    \addConstraint{\eqref{eq:sdpc1}-\eqref{eq:sdpc3}, \eqref{eq:sdpc5}, \eqref{eq:sdpc7}, \eqref{eq:minc1},  \eqref{eq:etaminusbetack},
    }{} \nonumber
    \addConstraint{ \eqref{eq:etaminusbetak},  \eqref{eq:ebb1ck}, \eqref{eq:ebb1k}, \eqref{eq:tauck}, \eqref{eq:tauk}, \eqref{eq:socck}, \eqref{eq:sock}, \eqref{eq:crbschur},
    }{} \nonumber
\end{maxi!}
where $\xi$ serves as a pre-defined penalty factor to minimize the penalty function to the greatest extent possible. However, since the penalty function exists, the objective function \eqref{eq:op4} is non-concave. Next, an iterative approach is employed to handle this challenge, leveraging the inequality given below by taking $\mathbf{P}_c$ as an example
\begin{equation}
\label{eq:rankonegeq}
\operatorname{tr}\left(\mathbf{P}_c\right)\!-\!\left(\mathbf{m}_{c, \mathsf{max}}^{(t)}\right)^H \mathbf{P}_c \mathbf{m}_{c, \mathsf{max} }^{(t)} \! \geq \! \operatorname{tr}\left(\mathbf{P}_c\right)\!-\!\chi_\mathsf{max}\left(\mathbf{P}_c\right) \!\geq\! 0,
\end{equation}
where $\mathbf{m}_{c, \mathsf{max}}^{(t)}$ is the normalized eigenvector at the $t^{\text{th}}$ optimization iteration with respect to the maximum eigenvalue $\xi_\mathsf{max}\left(\mathbf{P}_c\right)$. Same to this definition, we define $\mathbf{m}_{r, \mathsf{max}}^{(t)}$, $\mathbf{m}_{k, \mathsf{max}}^{(t)}$ as the normalized eigenvectors with regard to $\mathbf{P}_c$ and $\mathbf{P}_k$, respectively. Finally, the approximate problem at the $t^{\text{th}}$ iteration is formulated in \eqref{eq:p5},
\begin{figure*}[tb]
\begin{maxi!}|s|[2]                   % mini! = minimize  
    {\substack{\mathbf{P}_c, \mathbf{P}_r, \mathbf{c}, \{\mathbf{P}_k\}_{k \in \mathcal{K}}, \mu } }
    % optimization variable
    {\mu - \xi \mathsf{PF}
    \label{eq:op5}}  % objective function and label
    {\label{eq:p5}}             % label for opt problem
    % {\mathcal{P}1:} 
    % optimization result
    {} 
    \addConstraint{\eqref{eq:sdpc1}-\eqref{eq:sdpc3}, \eqref{eq:sdpc5}, \eqref{eq:sdpc7}, \eqref{eq:minc1},  \eqref{eq:etaminusbetack},\eqref{eq:etaminusbetak},  \eqref{eq:ebb1ck}, \eqref{eq:ebb1k}, \eqref{eq:tauck}, \eqref{eq:tauk}, \eqref{eq:socck}, \eqref{eq:sock}, \eqref{eq:crbschur}.
    }{} \nonumber
\end{maxi!}
\hrulefill
\end{figure*}
% \begin{figure*}[tb]
% \begin{maxi!}|s|[2]                   % mini! = minimize  
%     {\substack{\mathbf{P}_c, \mathbf{P}_r, \mathbf{c}, \{\mathbf{P}_k\}_{k \in \mathcal{K}}, \mu } }
%     % optimization variable
%     {\mu - \xi \bigl( [\operatorname{tr}\left(\mathbf{P}_c\right)-\left(\mathbf{m}_{c, \mathsf{max}}^{(t)}\right)^H \mathbf{P}_c \mathbf{m}_{c, \mathsf{max} }^{(t)} ]
%     + [\operatorname{tr}\left(\mathbf{P}_r\right)-\left(\mathbf{m}_{r, \mathsf{max}}^{(t)}\right)^H \mathbf{P}_r \mathbf{m}_{r, \mathsf{max} }^{(t)}]
%     \label{eq:op5}}  % objective function and label
%     {\label{eq:p5}}             % label for opt problem
%     % {\mathcal{P}1:} 
%     % optimization result
%     {} 
%     \breakObjective{+ \sum_{k \in \mathcal{K}} [\operatorname{tr}\left(\mathbf{P}_k\right)-\left(\mathbf{m}_{k, \mathsf{max}}^{(t)}\right)^H \mathbf{P}_k \mathbf{m}_{k, \mathsf{max} }^{(t)}] \bigl)} \nonumber
%     \addConstraint{\eqref{eq:sdpc1}-\eqref{eq:sdpc3}, \eqref{eq:sdpc5}, \eqref{eq:sdpc7}, \eqref{eq:minc1},  \eqref{eq:etaminusbetack},\eqref{eq:etaminusbetak},  \eqref{eq:ebb1ck}, \eqref{eq:ebb1k}, \eqref{eq:tauck}, \eqref{eq:tauk}, \eqref{eq:socck}, \eqref{eq:sock}, \eqref{eq:crbschur},
%     }{} \nonumber
% \end{maxi!}
% \hrulefill
% \end{figure*}
where $\mathsf{PF}$ is defined by
\begin{equation}
\begin{aligned}
    \mathsf{PF} & \triangleq [\operatorname{tr}\left(\mathbf{P}_c\right)-\left(\mathbf{m}_{c, \mathsf{max}}^{(t)}\right)^H \mathbf{P}_c \mathbf{m}_{c, \mathsf{max} }^{(t)} ]\\
    & \quad +\psi [\operatorname{tr}\left(\mathbf{P}_r\right)-\left(\mathbf{m}_{r, \mathsf{max}}^{(t)}\right)^H \mathbf{P}_r \mathbf{m}_{r, \mathsf{max} }^{(t)}] \\
    & \quad + \sum_{k \in \mathcal{K}} [\operatorname{tr}\left(\mathbf{P}_k\right)-\left(\mathbf{m}_{k, \mathsf{max}}^{(t)}\right)^H \mathbf{P}_k \mathbf{m}_{k, \mathsf{max} }^{(t)}].
\end{aligned}
\end{equation}
% \begin{figure*}
% \begin{maxi!}|s|[2]                   % mini! = minimize  
%     {\substack{\mathbf{P}_c, \mathbf{P}_r, \mathbf{c}, \{\mathbf{P}_k\}_{k \in \mathcal{K}}, \mu } }
%     % optimization variable
%     {\mu - \xi \bigl( [\operatorname{tr}\left(\mathbf{P}_c\right)-\left(\mathbf{m}_{c, \mathsf{max}}^{(t)}\right)^H \mathbf{P}_c \mathbf{m}_{c, \mathsf{max} }^{(t)}] 
%     \label{eq:op5}}  % objective function and label
%     {\label{eq:p5}}             % label for opt problem
%     % {\mathcal{P}1:} 
%     % optimization result
%     {}  \nonumber
%     \breakObjective{+ [\operatorname{tr}\left(\mathbf{P}_r\right)-\left(\mathbf{m}_{r, \mathsf{max}}^{(t)}\right)^H \mathbf{P}_r \mathbf{m}_{r, \mathsf{max} }^{(t)} ]} 
%     \breakObjective{+ \sum_{k \in \mathcal{K}} [\operatorname{tr}\left(\mathbf{P}_k\right)-\left(\mathbf{m}_{k, \mathsf{max}}^{(t)}\right)^H \mathbf{P}_k \mathbf{m}_{k, \mathsf{max} }^{(t)}] \bigl)} \nonumber
%     \addConstraint{\eqref{eq:sdpc1}-\eqref{eq:sdpc3}, \eqref{eq:sdpc5}, \eqref{eq:sdpc7}, \eqref{eq:minc1},  \eqref{eq:etaminusbetack},
%     }{} \nonumber
%     \addConstraint{ \eqref{eq:etaminusbetak},  \eqref{eq:ebb1ck}, \eqref{eq:ebb1k}, \eqref{eq:tauck}, \eqref{eq:tauk}, \eqref{eq:socck}, \eqref{eq:sock}, \eqref{eq:crbschur},
%     }{} \nonumber
% \end{maxi!}
% \end{figure*}

The transformed problem is convex and only involves SOC constraints, and LMI, thus it can be efficiently solved via numerical tools, \eg CVX \cite{grant2009cvx}. In each iteration, a feasible solution is given around the previous solution. We summarize our proposed algorithm for addressing \eqref{eq:p5} in Algorithm \ref{alg:alg1}, where $\varepsilon$ is the iteration tolerance coefficient, $t_\mathsf{max}$ is the maximum iteration number. After iteration ends, an optimal solution is given, and eigenvalue decomposition (EVD) is used to obtain the optimized beamforming vectors, $\mathbf{p}_c, \mathbf{p}_r, \{\mathbf{p}_k\}_{k \in \mathcal{K}}$, from the SDP expressions.
\begin{algorithm}[t]
	\caption{Proposed Algorithm for Max-min Fairness Beamforming Design}
	\label{alg:alg1}
	\KwIn{$\mathbf{P}_c^{(t)}, \mathbf{P}_r^{(t)}, \left\{\mathbf{P}_k^{(t)}\right\}_{k \in \mathcal{K}}, \mathbf{H}, \mathbf{H}_r, P_t$.}  
	\KwOut{$\mathbf{P}_c^{*}, \mathbf{P}_r^*, \left\{\mathbf{P}_k^* \right\}_{k \in \mathcal{K}}, \lambda^*, \eta^*, \beta^*, \tau^*$.} 
	\BlankLine
	Initialize $t \leftarrow 0, \lambda^{(t)} = 0, \xi=0.5 \times 10^{-2},  \varepsilon = 10^{-3}, t_\mathsf{max}=100$, $\mathbf{P}_c^{(t)}, \mathbf{P}_r^{(t)}, \left\{\mathbf{P}_k^{(t)}\right\}_{k \in \mathcal{K}}$ by SVD;
	
	\While{\textnormal{no convergence of objective function \eqref{eq:op5} \textbf{\&}} $\quad t<t_\mathsf{max}$ }{
        $t \leftarrow t+1$.\\
        Solve problem \eqref{eq:p5} to obtain $\mathbf{P}_c^{(t)}, \mathbf{P}_r^{(t)}, \left\{\mathbf{P}_k^{(t)}\right\}_{k \in \mathcal{K}}$.\\
		Update $\lambda^{(t)} = \min\limits_{k \in \mathcal{K}} \quad \frac{\widetilde{R}_{\mathsf{total},k}(\mathbf{P}_r^{(t-1)}, \mathbf{P}_k^{(t-1)}, \boldsymbol{\Delta}, C_k^{(t-1)}, \psi)}{P_{\mathsf{chain}, k}(\mathbf{P}_k^{(t-1)}, \boldsymbol{\Delta})}$.
	}	
    Set $\lambda^* = \lambda^{(t)}$.\\
	Return $\mathbf{P}_c^{*}, \mathbf{P}_r^*, \left\{\mathbf{P}_k^* \right\}_{k \in \mathcal{K}}, \lambda^*, \eta^*, \beta^*, \tau^*$.
\end{algorithm}

\bpara{Computational Complexity and Convergence Analysis.}
The formulated problem in \eqref{eq:p5} involves SDP, SOC and LMI constraints, which can be efficiently solved by standard interior-point method (IPM) \cite{ye2011interior}. The computational complexity is given by $\mathcal{O}\bigl(\sqrt{n} \log(\varepsilon^{-1}) (mn^3 + m^2 n^2 + m^3) \bigl)$, where $\varepsilon$ is the convergence tolerance, $n$ is the size of the positive semi-definite matrix, $m$ is the number of the SDP constraints. In our case \eqref{eq:sdp}, $m = 4K+3, n=N_t$, therefore the computational complexity of the proposed Algorithm \ref{alg:alg1} is $\mathcal{O}\bigl( \log(\varepsilon^{-1}) (N_t^{3.5} + 4 K N_t^{2.5}) \bigl)$.

The convergence of the proposed Algorithm \ref{alg:alg1} is also analyzed here. In each iteration, a feasible solution is given around the previous solution. In addition, the optimal solution $\{ \mathbf{P}_c^{(t)}, \mathbf{P}_r^{(t)}, \left\{\mathbf{P}_k^{(t)} \right\}_{k \in \mathcal{K}}, \lambda^{(t)}, \eta^{(t)}, \beta^{(t)}, \tau^{(t)} \}$ at the $t^{\text{th}}$ iteration is also a feasible solution of the $(t+1)^{\text{th}}$ iteration. Hence, the objective function \eqref{eq:op5} is non-decreasing after each iteration. Additionally, the objective function is bounded by lower bounds to satisfy the rank-one constraint \cite{boyd2004convex} as given in \eqref{eq:rankonegeq}, and thus guarantee a convergence. Consequently, the obtained feasible solution of \eqref{eq:p5} satisfies the Karush–Kuhn–Tucker (KKT) optimality conditions, and \eqref{eq:p4} and \eqref{eq:p5} have the same KKT conditions at convergence \cite{boyd2004convex, marks1978general}. Note that the global optimum is not guaranteed at the given feasible solution point.

\bpara{Total EE maximization.}
The total EE maximization problem in the considered system model with low-resolution DACs is used as a benchmark for comparison. The total EE is defined by the sum-rate over the total power consumption as given in \eqref{eq:fop6}. This problem can be solved by a similar algorithm as Algorithm \ref{alg:alg1} due to the summation is an operation preserving convexity \cite{boyd2004convex}.
\begin{maxi!}|s|[2]                   % mini! = minimize 
    % {\mathbf{p}_c, \mathbf{p}_r, \{\mathbf{p}_k\}_{k \in \mathcal{K}}, \mathbf{c}}       
    {\substack{\mathbf{p}_c, \mathbf{p}_r, \mathbf{c}\\ \{\mathbf{p}_k\}_{k \in \mathcal{K}}} }  
    % optimization variable
    % {\frac{\min_{k \in \mathcal{K}} (C_k + R_k)}{{\sum_{k\in \mathcal{K}} \text{tr} (\mathbf{p}_k\mathbf{p}^H_k)} } 
    { \frac{\sum_{k \in \mathcal{K}} (C_k + R_k)}{ \sum_{k \in \mathcal{K}} {\left(P_\mathsf{RF}+1/\kappa \,  (\norm{\boldsymbol{\Delta} \mathbf{p}_k}^2  + 1/K \norm{\boldsymbol{\Delta} \mathbf{p}_c}^2) \right)}} \label{eq:fop6}}  % objective function and label
    {\label{eq:fp6}}             % label for opt problem
    % {\mathcal{P}1:} 
    % optimization result
    {} 
    \addConstraint{\eqref{eq:fp1c1} - \eqref{eq:fp1c5}.}{} \nonumber
\end{maxi!}
Without introducing the auxiliary variable $\mu$ to handle max-min form as in \eqref{eq:minp}, the transformed problem is rewritten as
\begin{maxi!}|s|[2]                   % mini! = minimize  
    {\substack{\mathbf{P}_c, \mathbf{P}_r, \mathbf{c}, \\ \{\mathbf{P}_k\}_{k \in \mathcal{K}} } }
    % optimization variable
    {\!\sum_{k \in \mathcal{K}}{\!  \widetilde{R}_{\mathsf{total},k}(\mathbf{P}_r, \mathbf{P}_k, \boldsymbol{\Delta}, C_k, \psi)} 
     \label{eq:op7}}  % objective function and label
    {\label{eq:p7}}             % label for opt problem
    % {\mathcal{P}1:} 
    % optimization result
    {} \nonumber
    \breakObjective{\!- \!\lambda \bigl(\sum_{k \in \mathcal{K}} P_{\mathsf{chain}, k}(\mathbf{P}_k, \boldsymbol{\Delta}) \bigl) - \xi \mathsf{PF}} 
    \addConstraint{\eqref{eq:sdpc1}-\eqref{eq:sdpc3}, \eqref{eq:sdpc5}, \eqref{eq:sdpc7},  \eqref{eq:etaminusbetack},\eqref{eq:etaminusbetak},  
    }{} \nonumber
    \addConstraint{\eqref{eq:ebb1ck}, \eqref{eq:ebb1k}, \eqref{eq:tauck}, \eqref{eq:tauk}, \eqref{eq:socck}, \eqref{eq:sock}, \eqref{eq:crbschur},}{} \nonumber
\end{maxi!}
Therefore, we have the following Algorithm \ref{alg:alg2}.
\begin{algorithm}[t]
	\caption{Algorithm for Total EE Beamforming Design}
	\label{alg:alg2}
	\KwIn{$\mathbf{P}_c^{(t)}, \mathbf{P}_r^{(t)}, \left\{\mathbf{P}_k^{(t)}\right\}_{k \in \mathcal{K}}, \mathbf{H}, \mathbf{H}_r, P_t$.}  
	\KwOut{$\mathbf{P}_c^{*}, \mathbf{P}_r^*, \left\{\mathbf{P}_k^* \right\}_{k \in \mathcal{K}}, \lambda^*_\mathsf{total}, \eta^*, \beta^*, \tau^*$.} 
	\BlankLine
	Initialize $t \leftarrow 0, \lambda^{(t)}_\mathsf{total} = 0, \xi=0.5 \times 10^{-2},  \varepsilon = 10^{-3}, t_\mathsf{max}=100$, $\mathbf{P}_c^{(t)}, \mathbf{P}_r^{(t)}, \left\{\mathbf{P}_k^{(t)}\right\}_{k \in \mathcal{K}}$ by SVD;
	
	\While{\textnormal{no convergence of objective function \eqref{eq:op7} \textbf{\&}} $\quad t<t_\mathsf{max}$ }{
        $t \leftarrow t+1$.\\
        Solve problem \eqref{eq:p7} to obtain $\mathbf{P}_c^{(t)}, \mathbf{P}_r^{(t)}, \left\{\mathbf{P}_k^{(t)}\right\}_{k \in \mathcal{K}}$.\\
		Update $\lambda^{(t)} = \frac{\sum_{k \in \mathcal{K}} \widetilde{R}_{\mathsf{total},k}(\mathbf{P}_r^{(t-1)}, \mathbf{P}_k^{(t-1)}, \boldsymbol{\Delta}, C_k^{(t-1)}, \psi)}{\sum_{k \in \mathcal{K}} P_{\mathsf{chain}, k}(\mathbf{P}_k^{(t-1)}, \boldsymbol{\Delta})}$.
	}	
    Set $\lambda^*_\mathsf{total} = \lambda^{(t)}$.\\
	Return $\mathbf{P}_c^{*}, \mathbf{P}_r^*, \left\{\mathbf{P}_k^* \right\}_{k \in \mathcal{K}}, \lambda^*_\mathsf{total}, \eta^*, \beta^*, \tau^*$.
\end{algorithm}
Subsequently, we select the minimal EE among all the users to be the baseline.
\begin{remark}
 {The ULA in the LEO-ISAC satellite system can be easily extended to uniform planar array (UPA) setup. If we assume that the transmit and receive UPAs on the LEO satellite are equipped $M_t$ and $M_r$ antenna elements, respectively. Azimuth and elevation angles of the target regarding the satellite UPAs are denoted by  $\theta_h$ and $\theta_v$, respectively. Thus, the radar echo model in \eqref{eq:radarsignal} can be rewritten as $\mathbf{Z}[\ell] = \alpha e^{\jmath 2 \pi f_d \ell T} \mathbf{b}(\theta_h, \theta_v) \mathbf{a}^\top(\theta_h, \theta_v) \mathbf{x}_\mathsf{q}[\ell] + \mathbf{N}_r[\ell]$. Here,  $\mathbf{a}(\theta_h, \theta_v) \in \mathbb{C}^{M_t \times 1}$ and $\mathbf{b}(\theta_h, \theta_v) \in \mathbb{C}^{M_r \times 1}$ denote transmit and receive steering vectors, respectively, given by $\mathbf{a}(\theta_h, \theta_v) = \exp(-\jmath [\mathbf{r}_X, \mathbf{r}_Y, \mathbf{r}_Z] \mathbf{k}(\theta_h, \theta_v) )$, and $\mathbf{b}(\theta_h, \theta_v) = \exp(-\jmath [\overline{\mathbf{r}}_X, \overline{\mathbf{r}}_Y, \overline{\mathbf{r}}_Z] \mathbf{k}(\theta_h, \theta_v) )$, where $\mathbf{r}_Z = \overline{\mathbf{r}}_Z = \mathbf{0}$ if the UPAs are deployed within $(X,Y)$ plane. Additionally, $\mathbf{k}\left(\theta_h, \theta_v\right)=\frac{2 \pi}{\lambda_c}\left[\cos \theta_h \cos \theta_v, \sin \theta_h \cos \theta_v, \sin \theta_v\right]^\top$. Subsequently, we can obtain the FIM and CRB on the azimuth and elevation angles, and utilize a similar optimization algorithm to design the system.
}
\end{remark}

\section{Numerical evaluation}
The objective of this section is to illustrate the superiority of the RSMA-assisted ISAC-LEO scheme using the proposed algorithm. We start from the investigation of the convergence behaviours. Next, we solve the optimization problem as in \eqref{eq:p5} to obtain the optimized precoders. We compare the communications and sensing performances of the following schemes: max-min fairness EE optimization: (a) RSMA with radar sequence with SIC, (b) RSMA with radar sequence without SIC, (c) RSMA without radar sequence, the schemes of total EE optimization: (d) RSMA with radar sequence with SIC, (e) RSMA with radar sequence without SIC, (f) SDMA, (g) OMA. Note that the SDMA's optimization problem can be solved via Algorithm \ref{alg:alg2} by turning off the common stream. {Additionally, OMA is a sub-strategy of SDMA by serving only one user in an orthogonal resource, which is optimized by Algorithm \ref{alg:alg2}.}

\subsection{Simulation Environment}
We consider an overloaded ($N_t < K$) RSMA-assisted ISAC-LEO satellite system, where the LEO satellite is equipped $N_t = 4$ transmit antennas and $N_r = 4$ receive antennas. We assume that $K=5$ users are in the coverage area; the target is located at $\theta=0^\circ$ with distance $r = 2000$ m; the relative velocity between LEO satellite and target in LOS direction is $v = 10$ m/s. Note that no user grouping is considered in the simulation. The transmit power budget is $P_t = 20$ dBm, and we normalize the noise power by  $\iota T_\mathsf{sys}B$, \ie $\sigma_k^2 = 1, \forall k \in \mathcal{K}$  \cite{yin2022rate}. 
% the thermal noise power at each user is $\sigma_k = 0$  dBm, $\forall k \in \mathcal{K}$. 
For the DAC power consumption model, we set $P_\mathsf{LP} = 14 \, \text{mW}, P_\mathsf{M} =0.3 \, \text{mW}, P_\mathsf{LO}=22.5 \, \text{mW}, P_\mathsf{H}=3 \, \text{mW}$ and $\kappa = 0.27$ \cite{ribeiro2018energy}. The simulation parameters of LEO satellite channel are given by carrier frequency, $f_c = 20$ GHz, satellite height {$d_\mathsf{sat} = 600$ km}, bandwidth $B = 25$ MHz, $3$  dB angle equaling to $0.4^\circ$, satellite antenna gain, $G_u = 17$  dBi, Boltzman's constant, $\iota = 1.38 \times 10^{-23}$ J/m, system noise temperature $T_\mathsf{sys} = 517$K, and rain fading parameters, $(\mu_\mathsf{rain}, \sigma^2_\mathsf{rain}) = (-2.6, 1.63)$. In the constraint, QoS requirement for each user, and CRB threshold are set as, $R_\mathsf{th} = 1$ and { $\rho = 10^{-2}$}, respectively. As for the quantized DACs, we assume that all RF chains are equipped with the same quantization bit DACs for simplicity, \ie $b_m = b, \forall m \in \mathbb{Z}$. We adopt the values for the quantization loss $\varrho$ when $b \leq 5$ as in \cite{Fan_J_2015}. When $b > 5$, we calculate $\varrho$ as given in \eqref{eq:delta}. The number CPI is $L = 1024$, and the signals, $\mathbf{s}$, consist of uncoded i.i.d zero mean, unit-energy quadrature phase shift keying (QPSK) symbols. We conduct $100$ Monte Carlo simulations with perfect CSI to observe the performances.

Starting from the convergence analysis, \fig{fig:converg} (a) and (b) respectively demonstrate the changes of max-min fairness EE values and total EE values with respect to the iterations. The proposed algorithm can converge fast in $10$ iterations for all cases. Specifically, RSMA with $\mathbf{p}_r$ and SIC at the receiver with metric max-min fairness EE has the best performances in the optimization process via Algorithm \ref{alg:alg1}. In contrast, RSMA with $\mathbf{p}_r$ and no SIC has the lowest max-min fairness EE due to the interference from the radar sequence.
\begin{figure}[tb]
	\begin{center}
		\includegraphics[width =0.48\textwidth]{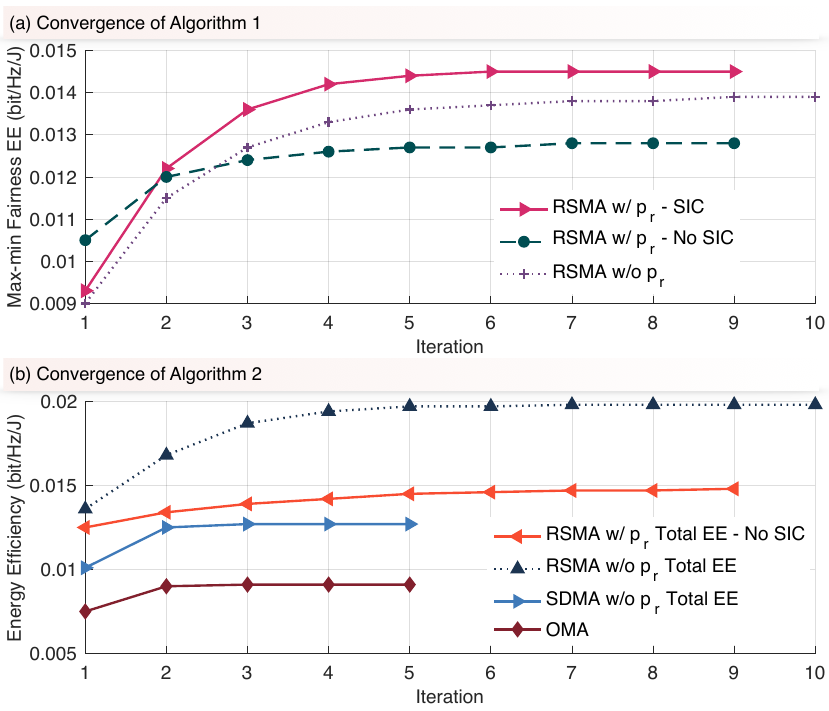}
		\caption{{Convergence illustration for 2-bit cases (a) Algorithm \ref{alg:alg1}, and (b) Algorithm \ref{alg:alg2} with $20$ dB transmit SNR and $40$ dB radar SNR.}}
		\label{fig:converg}
	\end{center}
\end{figure}

% \fig{fig:converg} shows
% \begin{itemize}
%     \item The proposed algorithm can converge fast in $15$ iterations.
%     \item RSMA with $\mathbf{p}_r$ with metric max-min fairness EE has the best performances in the optimization process. In contrast, SDMA has the worst ones.
% \end{itemize}

\subsection{Communication Performances}
We investigate the max-min fairness EE of the system as shown in \fig{fig:eevssnr} under the case that quantization bit, $b \in \{2, 4, 8\}$, and QoS requirements for each user, $R_\mathsf{th} = 1$. The results are obtained by solving \eqref{eq:p5} for the max-min fairness EE and \eqref{eq:p7} for the maximization of total EE. {OMA is a sub-case of SDMA, where only one user at one orthogonal resource is scheduled \cite{schroder2023comparison}. Thus, there is no inter-user interference in the SINRs \eqref{eq:sinr1} and \eqref{eq:sinr2}. Algorithm \ref{alg:alg2} is utilized to maximize this minimal EE of the OMA case.}
According to the \fig{fig:eevssnr}, the observations are that the max-min fairness EE increases with the decrease of the quantization bit, especially in the high transmit SNR range (\ie $20-30$  dB). The reason behind is due to a lower power consumption caused by fewer quantization bit. Additionally, RSMA with $\mathbf{p}_r$  and SIC at the receiver has the largest max-min fairness EE. {In contrast, OMA has the lowest one due to its lower spectrum efficiency compared to RSMA and SDMA. RSMA outperforms SDMA, because RSMA has a better inter-beam interference management.} In the five RSMA scenarios, RSMA with $\mathbf{p}_r$ and SIC has the best max-min EE performance, followed by RSMA without $\mathbf{p}_r$ and RSMA with $\mathbf{p}_r$ and no SIC. This is because that RSMA with $\mathbf{p}_r$ and SIC has no radar sequence interference (\ie $\psi = 0$) compared with RSMA with $\mathbf{p}_r$ and no SIC (\ie $\psi = 1$). In addition, RSMA with $\mathbf{p}_r$ and SIC has a better communication-sensing trade-off, enabled by the precoder, $\mathbf{p}_r$, for sensing, when it is compared with other cases. The max-min fairness EE of RSMA without $\mathbf{p}_r$ is higher than that of RSMA with $\mathbf{p}_r$ without SIC due to no radar sequence interference (\ie $\psi = 0$). The max-min fairness EE performances of the two cases, which are the maximization of the total EE with $\mathbf{p}_r$ and without $\mathbf{p}_r$, are worse than the three cases of directly maximizing minimum EE. In detail, the performance of total EE without $\mathbf{p}_r$ is better than total EE with $\mathbf{p}_r$ in terms of the max-min fairness EE. This is also because the latter case has no radar sequence interference. 
\label{sec:result}
\begin{figure}[tb]
	\begin{center}
		\includegraphics[width =0.48\textwidth]{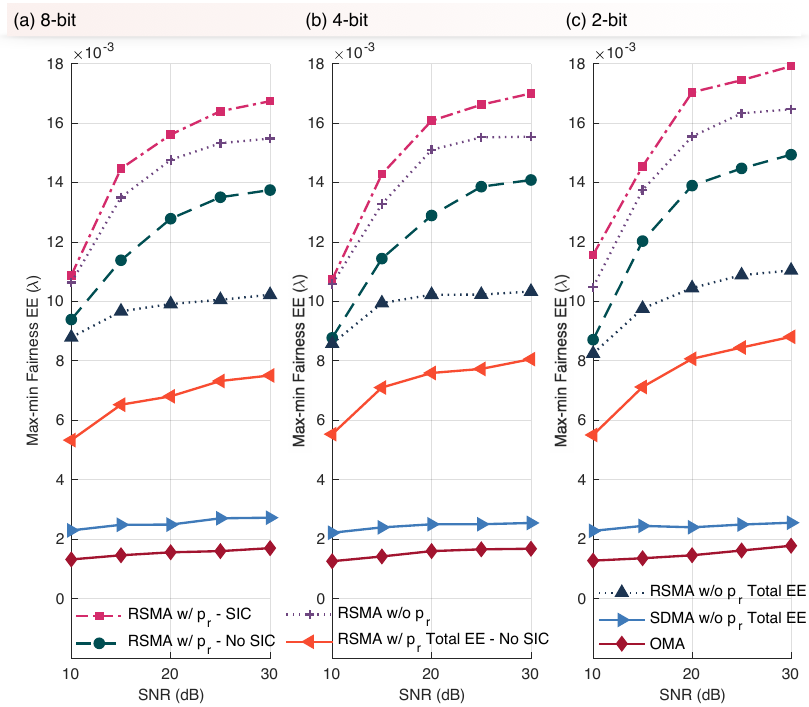}
		\caption{{Max-min fairness EE versus transmit SNR with the quantization bit $b \in \{2,4,8\}$ and $40$ dB radar SNR.}}
		\label{fig:eevssnr}
	\end{center}
\end{figure}

% \fig{fig:eevssnr} shows
% \begin{itemize}
%     \item With the decrease of the quantization bit, the energy efficiency increases.
%     \item RSMA with $\mathbf{p}_r$ has the largest max-min fairness energy efficiency. In contrast, SDMA has lowest one.
% \end{itemize}

We analyze the precoder power allocation here. The results are obtained by solving \eqref{eq:p5} and \eqref{eq:p7}. 
{As shown in \fig{fig:precoder_power} (a), the power allocation of each precoder of our proposed method is plotted. It is obvious that precoders $\mathbf{p}_r$ are effectively utilized under the max-min fairness EE metric. This utilization is further supported by the beampattern results shown in \fig{fig:angledetect}. This allows precoder $\mathbf{p}_c$ to focus on improving communication performance, specifically maximizing minimal EE. However, precoder $\mathbf{p}_r$ is not well used when adopting sum-rate metric (cf. \fig{fig:precoder_power} (e). Additionally, $\mathbf{p}_c$  serves both sensing and communication purposes, aligning with previous RSMA-ISAC research \cite{xu2021rate}. This dual-purpose usage restricts the potential of $\mathbf{p}_c$ to enhance communication performance.
}
% As shown in \fig{fig:precoder_power} (a), precoders $\mathbf{p}_c$ and $\mathbf{p}_r$ are well used in our proposed method that RSMA with $\mathbf{p}_r$ by maximizing the max-min fairness EE with guaranteed QoS for each user. As a result, the communication (\ie max-min fairness EE) and sensing performances are balanced. In comparison, the power allocation of precoder $\mathbf{p}_r$ is almost zero. It means that $\mathbf{p}_r$ is not well utilized in the strategy that maximizes the sum-rate. 
This conclusion is also verified by beampattern performances of the max-min fairness EE and sum-rate metrics in \fig{fig:angledetect}.
% precoder $\mathbf{p}_c$ is not well utilized in the strategy that maximization of the total EE as shown in \fig{fig:precoder_power} (a)-4, and (b)-4. This is because, with the smallest possible total transmitted power, the rate-splitting scheme tends to close the common stream and allocates power to the private stream to increase EE. Consequently, the max-min fairness EE performances of such strategy deteriorate. By turning off the constraint \eqref{eq:fp1c5}, we observe the power allocation of precoders. Since no sensing functionality is needed, the precoder for radar use, $\mathbf{p}_r$, is closed as demonstrated in \fig{fig:precoder_power} (b)-1, (b)-2 and (b)-4. 
To further explain the reason why $\mathbf{p}_r$ is effective under the metric of max-min fairness EE, we compare the power allocations of this metric and sum-rate metric. Based on the optimization problem \eqref{eq:fp1}, we change the objective function to sum-rate, which is the numerator in  \eqref{eq:fop1} without minimization. Subsequently, we follow the same problem and constraints transformation, and obtain the power allocation of sum-rate metric. It is observed that the sum-rate metric tends to turn off the radar sequence, thus $\mathbf{p}_c$ will work for both communication and sensing functionalities, and be probed at the target direction. In contrast, the metric that is max-min fairness EE can utilize both $\mathbf{p}_c$ and $\mathbf{p}_r$, resulting in a better trade-off between communication and sensing capabilities.
% \begin{figure}[tb]
% 	\begin{center}
% 		\includegraphics[width =0.5\textwidth]{figure/power_allo_1017.pdf}
% 		\caption{Power allocation of each precoder with 8 and 2-bit quantization and $20$ dB transmit SNR and $40$ dB radar SNR (a)-1, (b)-1: RSMA with $\mathbf{p}_r$ and SIC, (a)-2, (b)-2: RSMA with $\mathbf{p}_r$ but no SIC, (a)-3, (b)-3: RSMA without $\mathbf{p}_r$, (a)-4, (b)-4: RSMA with $\mathbf{p}_r$ by maximizing total EE, (a)-5, (b)-5: RSMA without $\mathbf{p}_r$ by maximizing total EE, (a)-6, (b)-6: SDMA.}
% 		\label{fig:precoder_power}
% 	\end{center}
% \end{figure}
\begin{figure}[tb]
	\begin{center}
		\includegraphics[width =0.48\textwidth]{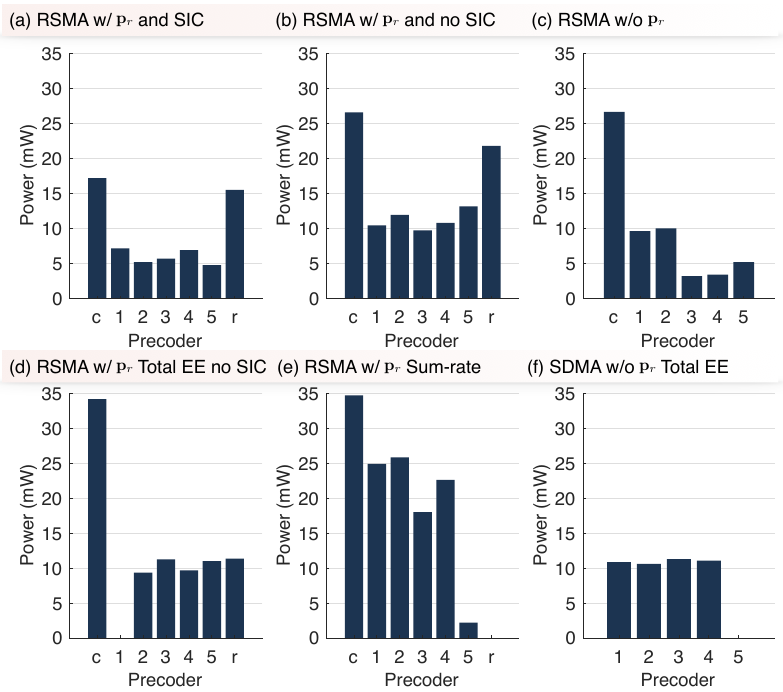}
		\caption{Power allocation of each precoder with  2-bit quantization and $20$ dB transmit SNR and $40$ dB radar SNR for (a) RSMA with $\mathbf{p}_r$ and SIC, (b) RSMA with $\mathbf{p}_r$ and no SIC, (c) RSMA without $\mathbf{p}_r$, (d) RSMA with $\mathbf{p}_r$ by maximizing total EE, (e) RSMA with $\mathbf{p}_r$ by maximizing sum-rate, (f) SDMA by maximizing total EE.}
		\label{fig:precoder_power}
	\end{center}
\end{figure}
In addition, as shown in \fig{fig:rcrbvssnr} (a), the max-min fairness EE decreases with $b$ owing to the increased power consumption.

\subsection{Sensing performances}
The sensing performance is constrained by $\mathsf{CRB}(\theta) \leq \rho$, where a smaller $\rho$ indicates a stronger constraint and results in a better sensing performance. As shown in \fig{fig:rcrbvssnr} (b), RSMA with $\mathbf{p}_r$ and SIC can achieve the best trade-off between the communication and sensing functionalities; the sensing constraint is satisfied while the highest max-min fairness EE can be achieved. In addition, the max-min fairness EE of each curve gradually increases with a looser $\rho$ constraint. 

% The rooted Cram\'er–Rao bound (RCRB) performances with respect to the SNR is shown in \fig{fig:rcrbvssnr}. The results are obtained by solving the optimization problems with the CRB constraint. From \fig{fig:rcrbvssnr}, RCRB increases when quantization bit decreases. This indicates that sensing capability decreases with the reduction of the quantization bit. SDMA has the worst RCRB, and other RSMA cases have the close RCRB performances, which indicates their sensing performances (\eg angle detection) are close.

% \begin{figure}[tb]
% 	\begin{center}
% 		\includegraphics[width =0.48\textwidth]{figure/rcrb1127.pdf}
% 		\caption{RCRB versus transmit SNR with the quantization bit $b={2,4,8}$.}
% 		\label{fig:rcrbvssnr}
% 	\end{center}
% \end{figure}
\begin{figure}[tb]
	\begin{center}
		\includegraphics[width =0.48\textwidth]{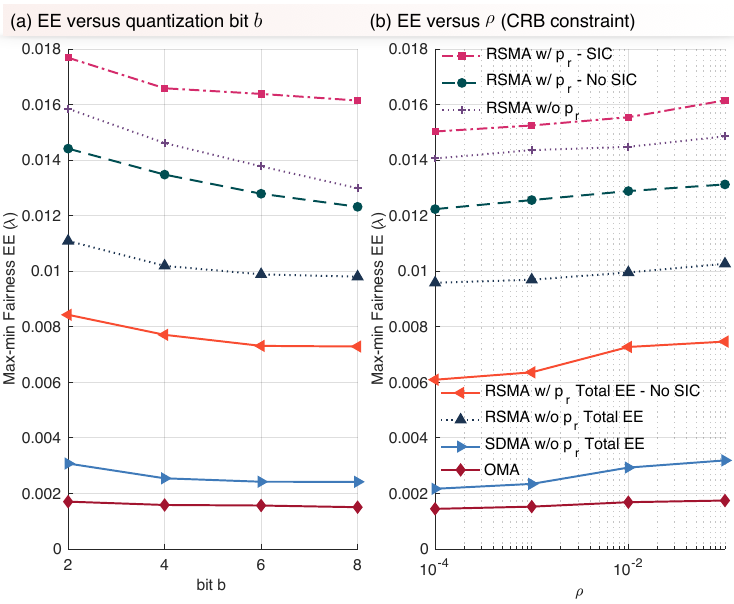}
		\caption{{Under the case that $20$ dB transmit SNR and $40$ dB radar SNR, (a) Max-min fairness EE versus bit $b$, (b) Max-min fairness EE versus RCRB constraint $\rho$ with $b=2$.}}
		\label{fig:rcrbvssnr}
	\end{center}
\end{figure}
% \fig{fig:rcrbvssnr} shows
% \begin{itemize}
%     \item With the decrease of the quantization bit, RCRB increases. This indicates that sensing capability decreases with the reduction of the quantization bit.
%     \item SDMA has the worst RCRB, and other RSMA cases have the close RCRB performances, which indicates their sensing performances (\eg angle, velocity, and range detection) are close. It can be validated later.
% \end{itemize}

The real performances of the sensing functionality are captured by the information extraction. Here, we investigate the range-Doppler detection and beampattern performances. We use the same sensing detection pipeline as in \cite{yin2022ratewcnc, liu2023joint}. Next, we perform a discrete Fourier transform (DFT) operation to extract the range-Doppler map based on the discrete-time signal stream in \eqref{eq:radarsignal}. According to \fig{fig:rd}, observations are that target's Doppler-range information can be extracted by solving the precoder optimization problems with CRB constraint.   {Specifically, RSMA with $\mathbf{p}_r$ and SIC has the highest value, \ie $\text{value} = 3.6026 \times 10^{-3}$ followed by RSMA with $\mathbf{p}_r$ and no SIC, and RSMA without $\mathbf{p}_r$. In contrast, SDMA has the smallest detection value. These results indicate the superiority of our proposed approach in terms of range-Doppler detection.}
\begin{figure}[tb]
	\begin{center}
		\includegraphics[width =0.48\textwidth]{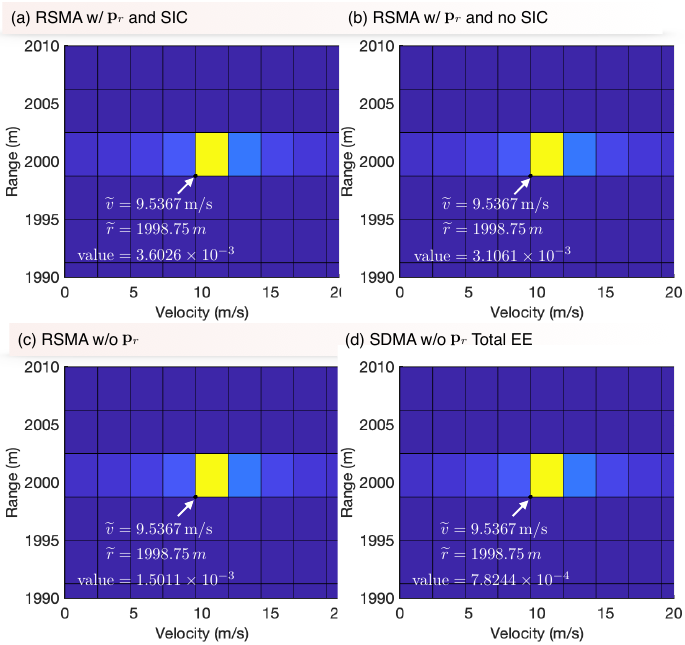}
		\caption{Detection results of Doppler and range with $2$-bit quantization, $20$ dB transmit SNR and $40$ dB radar SNR: Range-Doppler Maps for (a) RSMA with $\mathbf{p}_\text{r}$ and SIC, (b) RSMA with $\mathbf{p}_\text{r}$ and no SIC, (c) RSMA without $\mathbf{p}_\text{r}$, (d) SDMA by maximizing total EE.}
		\label{fig:rd}
	\end{center}
\end{figure}
To investigate how the precoders $\mathbf{p}_\text{c}$ and $\mathbf{p}_\text{r}$ are utilized in the two metrics (\ie max-min fairness EE and sum-rate), we plot the beampatterns under two scenarios. {The two scenarios are (1) users and target are distant (\ie all users are located around $-45^{\circ}$ and target is at $30^{\circ}$); (2) users and target are near (\ie all users are located around $-45^{\circ}$ and target is at $-10^{\circ}$).} As shown in \fig{fig:angledetect}, in both cases, $\mathbf{p}_\text{r}$ is probed at the target direction when maximizing max-min fairness EE. In contrast, $\mathbf{p}_\text{r}$ is not used when maximizing sum-rate. To achieve the sensing function, $\mathbf{p}_\text{c}$ of the sum-rate metric is probed at the target direction. This observation is consistent with previous RSMA ISAC studies \cite{xu2021rate}. Moreover, $\mathbf{p}_\text{c}$ of the max-min fairness EE metric is not completely probed at the target direction, and thus enables flexibility and a higher max-min fairness EE performance.

\begin{figure}[tb]
	\begin{center}
		\includegraphics[width =0.5\textwidth]{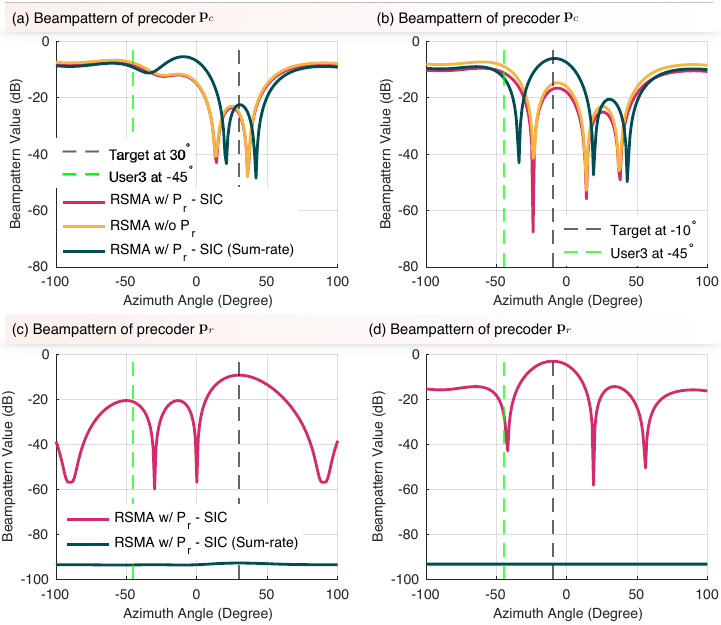}
		\caption{Under $2$-bit quantization, $20$ dB transmit SNR, and $40$ dB radar SNR. Case 1: users and targets are distant, transmit beampatterns of (a) common stream precoder $\mathbf{p}_\text{c}$, (c) radar sequence precoder $\mathbf{p}_\text{r}$. Case 2: users and targets are near, transmit beampatterns of (b) common stream precoder $\mathbf{p}_\text{c}$, (d) radar sequence precoder $\mathbf{p}_\text{r}$.}
		\label{fig:angledetect}
	\end{center}
\end{figure}
% \fig{fig:angledetect} shows
% \begin{itemize}
%     \item As shown in Fig. (a), the transmit beampattern of our approach can probe at the desired angles (\ie directions of target and the $4$ users).
%     \item The sum of all precoders $\mathbf{p}_\text{sum}$, $\mathbf{p}_\text{c}$, and $\mathbf{p}_\text{r}$ will probe at the $5$ desired directions.
%     \item Precoder $\mathbf{p}_1$ for the user 1 will probe the user 1's direction.
%     \item The incoming angles of the target and $4$ users can be well detected by the MUSIC algorithms.
%     \item The angle from the target has the highest values, and the angle from $4$ users have the relatively low values.
% \end{itemize}

\begin{remark}
{Compared to prior RSMA ISAC studies \cite{xu2021rate, yin2022ratewcnc, yin2022rateletter, dizdar2022energy,loli2022rate, chen2023rate}, the consistent observation is that when maximizing sum-rate, radar sequence will not be utilized, and the common stream will be used for both communication and sensing functions. Specifically, it will be probed at the target direction for sensing guarantee. The new observation is that when adopting max-min fairness EE as the metric, radar sequence can be utilized, resulting in the improvement of the minimal EE among users. The reason behind this is that the utilization of a separated radar sequence is more energy-efficient in terms of communication use. In detail, the power of radar sequence is used for sensing, and the power consumed by common and private streams, which determines EE, is minimized mainly for communication use.
}

\end{remark}

\section{Conclusion}
In this paper, we have explored an RSMA-assisted ISAC-LEO satellite system that incorporates the low-resolution DACs at the transmitter. This system serves multiple communication users and detects a moving target, simultaneously. We have investigated the max-min fairness EE of the proposed strategy under communication and sensing constraints. This problem has been formulated as a non-convex optimization problem by designing the precoders. An iterative algorithm based on SCA and Dinkelbach's method has been proposed to solve the problem. 
{Numerical results have demonstrated that the proposed design outperforms the strategies that maximize the total EE of the system, SDMA and OMA in terms of max-min fairness EE and communication-sensing trade-off. }

\section*{Appendix A: Derivation of The Fisher Information Matrices} 
The elements at the $i^{\text{th}}$ column and $j^{\text{th}}$ row in the FIM, $\mathbf{J}$, is given by \cite{Kay_B_1993, bekkerman2006target}
\begin{equation}
[\mathbf{J}]_{i, j}=\frac{2}{\sigma_r^2} \operatorname{Re}\left\{\sum_{\ell=1}^L \frac{\partial \boldsymbol{\mu}[\ell]^H}{\partial \xi_i} \frac{\partial \boldsymbol{\mu}[\ell]}{\partial \xi_j}\right\}, i, j \in \{1, 2\},
\end{equation}
where $\boldsymbol{\mu} \triangleq \alpha e^{\jmath 2 \pi f_d \ell T} \mathbf{A}(\theta) \boldsymbol{\Delta}  \mathbf{x}[\ell]$. The partial derivatives of $\boldsymbol{\mu}$ with respect to the unknown parameters are given by
\begin{align}
\frac{\partial \boldsymbol{\mu}}{\partial \theta}
& =\alpha e^{\jmath 2 \pi f_d \ell T} \dot{\mathbf{A}}(\theta) \boldsymbol{\Delta}  \mathbf{x}[\ell], \\
\frac{\partial \boldsymbol{\mu}}{\partial \alpha} 
& = (1, \jmath) \otimes e^{\jmath 2 \pi f_d \ell T} \mathbf{A}(\theta) \boldsymbol{\Delta}  \mathbf{x}[\ell].
\end{align}
Therefore, the expression of each element is given by
\begin{align}
    J_{\theta \theta}  & = \frac{2 L |\alpha|^2}{\sigma_r^2} \operatorname{tr}\left(\dot{\mathbf{A}}(\theta) \mathbf{R}_\mathsf{q}\dot{\mathbf{A}}^H(\theta)\right),\\
    J_{\theta \alpha} & = \frac{2 L |\alpha|^2}{\sigma_r^2} \operatorname{Re}\left\{\alpha^* \operatorname{tr}\left(\mathbf{A}\left(\theta_p\right) \mathbf{R}_\mathsf{q} \dot{\mathbf{A}}^H\left(\theta_l\right)\right)(1, j)\right\}, \\
    J_{\alpha \alpha} & \! = \! \frac{2 L |\alpha|^2}{\sigma_r^2} \operatorname{Re}\left\{(1, j)^H(1, j) \operatorname{tr}\left(\mathbf{A}(\theta) \mathbf{R}_\mathsf{q} \mathbf{A}^H(\theta)\right)\right\},
\end{align}
where the derivative of $\mathbf{A}(\theta)$ is given by
\begin{equation}
\label{eq:dA}
\begin{aligned}
    \dot{\mathbf{A}}(\theta) &= \frac{\partial \mathbf{A}(\theta)}{\partial \theta}=\frac{\partial \mathbf{b}(\theta)}{\partial \theta} \mathbf{a}^\top(\theta)+\mathbf{b}(\theta) \frac{\partial \mathbf{a}^\top(\theta)}{\partial \theta}\\
    & = \jmath2\pi\delta \cos{\theta} \big( \text{diag}(\Bar{\mathbf{r}}_{X}) \mathbf{b}(\theta) \mathbf{a}^\top(\theta) \\
    & \quad + \mathbf{b}(\theta) \mathbf{a}^\top(\theta) \text{diag}(\mathbf{r}_{X}) \big).
\end{aligned}
\end{equation}
In \eqref{eq:dA}, $\mathbf{r}_{X} = [0, \cdots, N_r-1]^\top$ and $\Bar{\mathbf{r}}_{X} = [0, \cdots, N_t-1]^\top$ represent the Cartesian coordinates of the transmit and receive antenna arrays.

\bibliographystyle{IEEEtran_url}
\bibliography{IEEEabrv,references,ABlist}

\end{document}